\documentclass[10pt,twocolumn,letterpaper,pagebackref,breaklinks,colorlinks,citecolor=cvprblue]{article}

\usepackage{cvpr}              

\usepackage{tabularx}
\usepackage{multirow}
\usepackage{adjustbox} 
\usepackage{enumitem}
\usepackage{subcaption}
\usepackage{comment}
\usepackage{xcolor}
\usepackage{colortbl}

\newif\ifarxiv
\arxivfalse

\definecolor{lightgray}{rgb}{0.88,0.88,0.88}

\newcommand{\stdhide}[1]{}

\newcommand{\red}[1]{\textcolor{red}{#1}}
\newcommand{\pink}[1]{\textcolor{pink}{#1}}

\newcommand{\floorm}{floormask\xspace}
\newcommand{\archm}{archmask\xspace}

\newcommand{\diffuscene}{\textsc{DiffuScene}\xspace}
\newcommand{\atiss}{\textsc{ATISS}\xspace}
\newcommand{\layoutgpt}{\textsc{LayoutGPT}\xspace}
\newcommand{\semdiff}{\textsc{SemLayoutDiff}\xspace}
\newcommand{\midiff}{\textsc{MiDiffusion}\xspace}

\newcommand{\fid}{\text{FID}$\downarrow$ \xspace}
\newcommand{\kid}{\text{KID}$\downarrow$ \xspace}
\newcommand{\sca}{\text{SCA}$\%$ \xspace}
\newcommand{\ckl}{\text{CKL}$\downarrow$ \xspace}

\newcommand{\semtype}[1]{\texttt{\small #1}\xspace}

\newcommand{\oobscene}{$\text{OOB}_S\! \downarrow$\xspace}
\newcommand{\oobobj}{$\text{OOB}_O\! \downarrow$\xspace}
\newcommand{\col}{\text{COL}$\downarrow$\xspace}
\newcommand{\nav}{\text{NAV}$\uparrow$\xspace}

\newcommand{\mypara}[1]{\noindent\textbf{#1}}

\newcommand\best[1]{\textbf{#1}}

\newcolumntype{Y}{>{\centering\arraybackslash}X}

\newcommand\imgclip[2]{\includegraphics[trim={#1 #1 #1 #1},clip]{#2}}

\expandafter\def\expandafter\normalsize\expandafter{%
    \normalsize%
    \setlength\abovedisplayskip{5pt}%
    \setlength\belowdisplayskip{7pt}%
    \setlength\abovedisplayshortskip{-8pt}%
    \setlength\belowdisplayshortskip{2pt}%
}

\definecolor{cvprblue}{rgb}{0.21,0.49,0.74}
\usepackage[pagebackref,breaklinks,colorlinks,citecolor=cvprblue]{hyperref}

\def\paperID{{215}} 
\def\confName{3DV\xspace}
\def\confYear{2026\xspace}

\newcommand{\figfirstpagefigure}{
\begin{center}
\includegraphics[width=\textwidth]{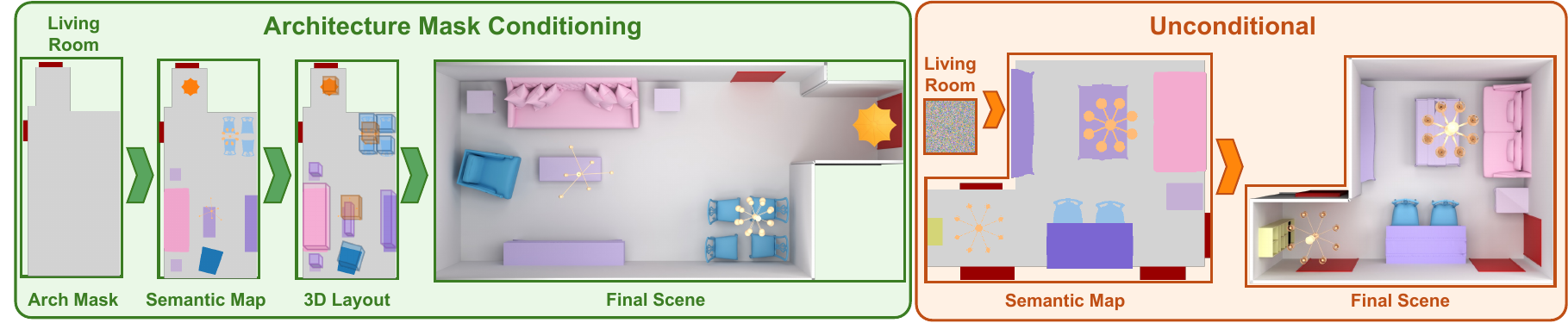}
\captionof{figure}{
\semdiff generates 3D scenes conditioned on an architectural map or unconditionally. Left (full pipeline): With architectural conditioning and room type label, \semdiff synthesizes a 2D semantic layout map, predicts 3D attributes to form bounding box layouts, and retrieves objects to construct a final scene. Right: In the unconditional setting, \semdiff generates the architecture map and the semantic layout map from noise and room type label before synthesizing the scene (3D layout stage not shown).
}
\label{fig:teaser}
\end{center}
}

\begin{document}

\title{SemLayoutDiff: Semantic Layout Generation with Diffusion Model\\for Indoor Scene Synthesis}

\author{
Xiaohao Sun$^1$ \quad Divyam Goel $^2$ \quad Angel X. Chang $^{1,3}$\\
$^1$Simon Fraser University \quad $^2$CMU \quad $^3$Alberta Machine Intelligence Institute (Amii)
}

\twocolumn[{
\maketitle
\vspace{-2.5em}
\centerline{\href{https://3dlg-hcvc.github.io/SemLayoutDiff/}{3dlg-hcvc.github.io/SemLayoutDiff/}}
\figfirstpagefigure
}]

\begin{abstract}
We present SemLayoutDiff, a unified model for synthesizing diverse 3D indoor scenes across multiple room types.
The model introduces a scene layout representation combining a top-down semantic map and attributes for each object. 
Unlike prior approaches, which cannot condition on architectural constraints, SemLayoutDiff employs a categorical diffusion model capable of conditioning scene synthesis explicitly on room masks. 
It first generates a coherent semantic map, followed by a cross-attention-based network to predict furniture placements that respect the synthesized layout.
Our method also accounts for architectural elements such as doors and windows, ensuring that generated furniture arrangements remain practical and unobstructed.
Experiments on the 3D-FRONT dataset show that SemLayoutDiff produces spatially coherent, realistic, and varied scenes, outperforming previous methods.
\end{abstract}
\vspace{-1em}
\section{Introduction}
\label{sec:intro}
Automatic 3D indoor environment generation has many applications such as assisting content designers for AR/VR, video games, and serving as training data for computer vision models~\cite{infinigen2023infinite} and embodied AI agents~\cite{deitke2022procthor}.
Various approaches have been proposed for generating 3D indoor scenes. 
Typically, these methods separate the task into two steps: 1) generating a coarse layout specifying the semantic categories and positions of objects, and 2) retrieving and placing suitable objects based on the generated layout.

Early attempts used design guidelines~\cite{merrell2011interactive} to determine the object arrangement.
These approaches used hand-crafted rules, limiting generalization to diverse types of scenes.
\citet{fisher2012example} introduced data-driven object placement.
Since then, various deep learning techniques have been used for layout generation, including autoregressive models~\cite{wang2018deep,wang2021sceneformer,paschalidou2021atiss}, graph neural networks~\cite{wang2019planit}, and diffusion over graphs~\cite{tang2023diffuscene}.
More recently, LLMs have been used for open-vocabulary scene generation~\cite{aguina2024open}.

While LLMs are useful for providing priors on what objects are present and semantic relations between objects, they struggle to precisely place objects.
Thus, researchers still actively investigate what can be learned via training on 3D scenes~\cite{tang2023diffuscene,hu2024mixed}.
Recent works trained on 3D scenes share common limitations: 1) room architecture is not handled, 2) interpenetration of objects, 3) lack of unified model that can be conditioned with different inputs (separate models typically trained for each room type).

To address the first two issues, we propose the use of a 2D top-down semantic map to represent the layout.
In this semantic map, each cell represents one object category.
By including architectural elements such as floors, doors, and windows, the representation also accommodates 
\emph{generation of the room} as well as \emph{layout of objects} in the room.
The representation naturally ensures that objects do not overlap and floorplan constraints are properly maintained as shown in related work~\cite{zhang2020deep,su2025chord}. 
Specifically, we use a categorical diffusion model to generate the top-down semantic map. 
From the semantic map, we extract the object instances and their attributes (e.g., semantic category, size, orientation). 

We also tackle training a unified model across room types that can be conditioned on the architecture and room type.  
We demonstrate that we can train a unified model that handles different room types and generates more plausible and realistic layouts.
In summary, our contributions are:
\begin{itemize}
    \item A novel scene representation using semantic layout maps with instance attributes, which enables simultaneous object placement and better captures spatial relationships 
    \item A categorical diffusion model for scene synthesis by generating semantic maps, enabling a unified approach that efficiently handles diverse layouts across room types, captures architectural constraints and furniture relationships, and reduces out-of-bound and object intersection issues.
    \item Our model incorporates architectural elements by generating the room together with the objects, and conditioning on the architecture. 
\end{itemize}

\section{Related work}
\label{sec:related}

\mypara{Rule-based and statistical prior indoor scene synthesis.}
Early work relied on placement constraints~\cite{xu2002constraint} and rules based on interior design principles~\cite{merrell2011interactive} to place objects.  Following these works, \citet{fisher2012example} learned object arrangements from a 3D scene database by modeling the co-occurrence and spatial relationship of pairs of objects and constructing scenes hierarchically based on support.

\mypara{Deep-learning based scene synthesis.}
\citet{wang2018deep}, introduced auto-regressive scene generation and used CNNs for scene synthesis by representing scenes as 2D top-down images, with multiple channels encoding information such as floor layout, object semantic mask, etc. 
Followup work improved generation efficiency~\cite{ritchie2019fast}, and used transformers to decode the objects~\cite{wang2021sceneformer,paschalidou2021atiss,para2022cofs,sun2024forest2seq}. 
Scenes were also modeled using graphs: a scene-hierarchy~\cite{li2019grains,gao2023scenehgn}, a relationship graph~\cite{wang2019planit,zhou2019scenegraphnet,luo2020end,zhai2023commonscenes}, or a hybrid representation that combines scene-graphs with top-down image based representation~\cite{zhang2020deep}.    
Recent work~\cite{tang2023diffuscene,lin2024instructscene,sun2025reltriple} trains denoising diffusion model to generate the scene graph, with some work incorporating a floor plan as a conditioning signal~\cite{hu2024mixed,maillard2024debara} and loss terms to avoid collisions~\cite{yang2024physcene}.

We advocate the use of 2D top-down images as in \citet{wang2018deep}.
Instead of autoregressively adding one object at a time, we use diffusion to layout all objects together.  By working directly in image space, the model can ensure that the position of the objects is within the floor plan.  
Concurrent to our work, \citet{su2025chord} used diffusion-based models to generate a semantic layout conditioned on floor plan, but uses continuous instead of discrete categorical outputs.

\mypara{LLM/VLM-based scene generation.} As LLMs and VLMs capture common-sense knowledge about object placements in rooms, researchers developed frameworks that leverage LLMs to generate more open-world scenes based on text~\cite{wen2023anyhome,yang2024holodeck,hu2024scenecraft,ccelen2024design} and/or visual information~\cite{sun2024layoutvlm}. 
While the use of LLMs/VLMs is a promising direction, we investigate whether we can train a unified model that generates reasonable semantic layouts using diffusion.

\mypara{Combined layout and object generation.} 
With advances in 3D shape generation, recent work has studied how to generate objects for a specified layout using NeRF in a compositional setting~\cite{cohen2023set,lin2023componerf,po2023compositional}.
Our model can be used to specify the initial layout used by these compositional approaches. 
Some works generate both the layout and the objects~\cite{vilesov2023cg3d,gao2023graphdreamer,zhai2023commonscenes,zhai2024echoscene}, typically with a graph-based representation for the layout and then a shape generator for each object. Some~\cite{zhai2023commonscenes,zhai2024echoscene} still learn priors on the placement of objects from 3D datasets such as 3D-FRONT~\cite{fu20213d}, while others~\cite{gao2023graphdreamer} use an LLM to convert text to semantic scene-graphs, which are then used to generate relative positions of objects.
We demonstrate in \cref{fig:vis-trellis} that layouts generated by our model can also be combined with generated objects.

\mypara{Diffusion models for layout generation.}
Diffusion has also been applied to 2D layout synthesis~\cite{sohl2015deep,ho2020denoising, song2020denoising, song2020improved}.
\citet{chai2023layoutdm} explore graphic layout generation with a newly designed transformer-based denoiser. \citet{inoue2023layoutdm} formulate a discrete diffusion model for layout generation, which can solve diverse tasks via a single model using complex layout constraints.
Another work~\cite{hoogeboom2021argmax} shows that using a multinomial diffusion model over categorical data, it is possible to generate 2D semantic maps.
We take inspiration from diffusion for 2D layouts, and use diffusion to generate a 2D top-down semantic map that captures the spatial relationship between all objects.
\section{Semantic Layout Representation}
\label{sec:representation}

In recent scene generation work, the scene is represented as a sequence of objects~\cite{paschalidou2021atiss} with corresponding attributes (category, location, etc.) or a scene graph~\cite{tang2023diffuscene} where nodes indicate objects and edges indicate object relations.
Both approaches face the problem of overlapping objects and do not necessarily respect the floor boundary.

\begin{figure}
\centering
\includegraphics[width=\linewidth]{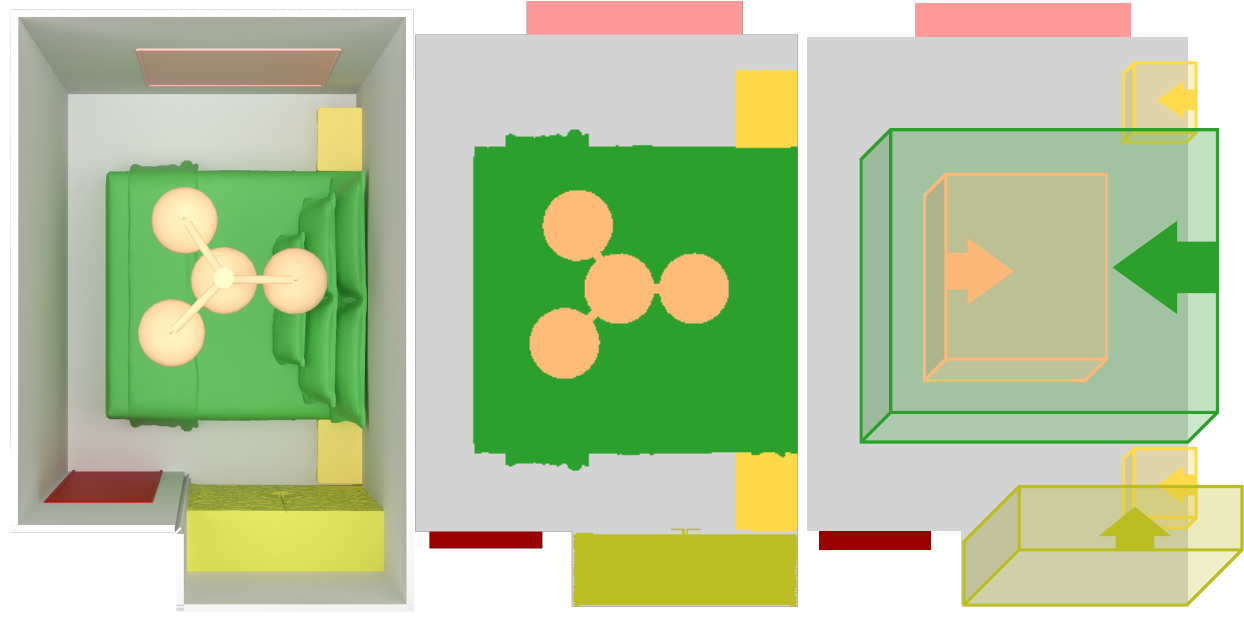}
\vspace{-18pt}
\caption{Semantic map representation example. 
We represent the scene in two parts: the 2D top‑down semantic map with a fixed physical unit per pixel (middle) and the 3D bounding boxes with orientations for object‑level attributes (right). Note that the left image is the corresponding rendered scene.
}
\vspace{-12pt}
\label{fig:data_example}
\end{figure}

To tackle this problem, we represent the scene as a semantic map with exact instance annotation, shown in \cref{fig:data_example}.
The semantic map, denoted as $\mathcal{S} \in \mathbb{R}^{H \times W}$, represents the semantic segmentation of a top-down projection of a room, encoding both location and object category. 
Each pixel corresponds to fixed physical dimensions.
We use a scale of $s=0.01$ meters (i.e., a pixel is 0.01 meters).
We denote the instance-level annotations as $\mathcal{I} = \{O_i\}$ where $i = 1, ..., N$, where $N$ is the total number of objects within the room. 
Each instance $O_i$ is defined by $O_i = \{c_i, s_i, p_i, r_i\}$, representing an object's category, size, position, and orientation.
The category $c_i \in \{0, 1\}^K$ is a categorical variable over the total number of semantic categories $K$ in the dataset, where $K = C + 4$. 
Here, $C$ is the number of object types, and the additional categories represent architectural elements: \semtype{void} (outside room boundaries), \semtype{floor}, \semtype{door}, and \semtype{window}. 
Including floors, doors, and windows enables our model to generate the room together with the objects, while also handling architectural constraints.
Our framework can be simplified to scenarios without doors and windows by reducing the semantic categories to $K = C + 2$, representing only \semtype{floor} and \semtype{void}.
This simplified setting is similar to conditioning by floor as in prior work~\cite{paschalidou2021atiss, hu2024mixed}.
However, our method simultaneously generates the room and object layouts, even under unconditional generation.

For each object instance $O_i$, in addition to the semantic category $c_i$, we also specify the bounding box size $s_i \in \mathbb{R}^3$,  position $p_i \in \mathbb{R}^3$, and orientation $r_i$. 
For the orientation, we only consider the rotation of objects about the up (vertical) axis.
From analyzing the orientation of objects in 3D-FRONT, we observed that over $97\%$ of orientations are aligned with the coordinate axes.
Thus we restrict our problem to predicting four distinct orientation classes and use a categorical variable $r_i \in \{0, 1, 2, 3\}$ to indicate the object's front direction corresponding to $0^{\circ}, 90^{\circ}, 180^{\circ}, 270^{\circ}$.

We process the scene data using BlenderProc~\cite{denninger2023blenderproc2} to render top-down views of the rooms, producing both semantic map and instance-level annotations. 
The camera parameters ensure consistent pixel-to-meter scaling. 
Images are padded to a fixed size ($1200 \times 1200$, representing $12	\text{m} \times 12 \text{m}$). 
Finally, we apply the filtering strategy from ATISS~\cite{paschalidou2021atiss} to ensure data quality. 
More details are in \cref{sec:supp-data-processing}.

\section{Method}

\begin{figure*}
\vspace{-1em}
\includegraphics[width=\textwidth]{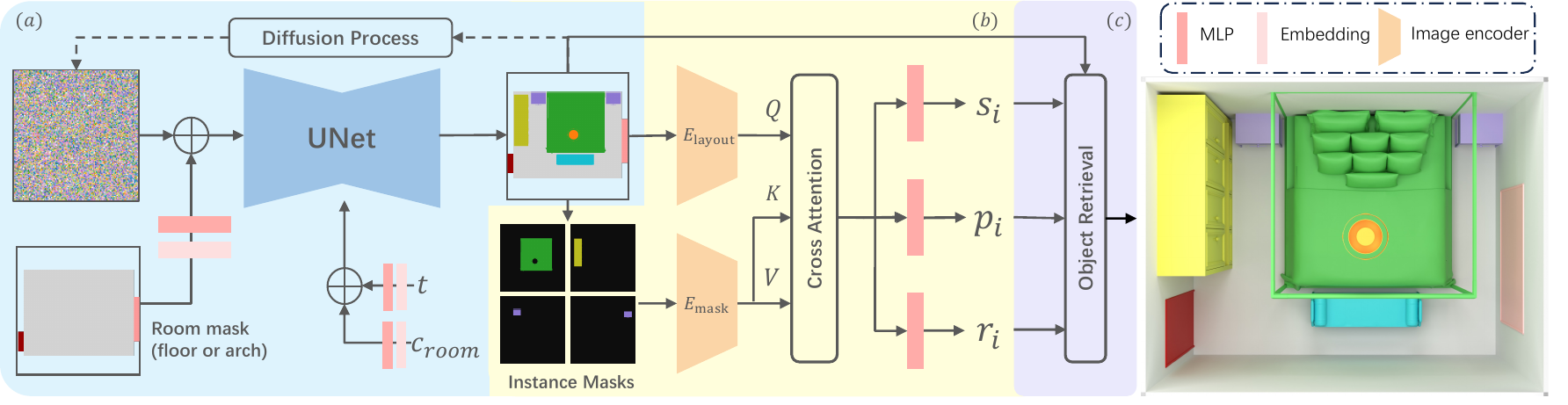}
\vspace{-2em}
\caption{\textbf{SemLayoutDiff overview}. From left: (a) is the unified diffusion model that is conditioned on the room mask, and room type $c_{room}$. During the denoising process, the \archm or \floorm embedding is added to the noise input embedding. The room type embedding is added to the timestep embedding. (b) is the object attribute prediction model with a semantic layout map as input. $s_i, p_i, r_i$ indicate the $i$th instance's size, position, and orientation. At training time, we use ground-truth instance masks. During inference the semantic layout map is split into instance masks by using connected component analysis. The layout feature and the mask feature are passed to a cross-attention layer to get the final object instance feature, which is used to predict attributes.
(c) During inference,  objects are retrieved to match the category $c_i$ and size $s_i$, and arranged using the position $p_i$ and orientation $r_i$. 
}
\vspace{-1em}
\label{fig:semdiff}
\end{figure*}

Unlike previous methods~\cite{paschalidou2021atiss, tang2023diffuscene, hu2024mixed}, our unified model generates all room types with a single model.
It generates both the architecture and furniture objects at the same time in an unconditional way.
Furthermore, it takes a \textit{room mask} as input condition, and generates layouts conditioned on just the floormask or the full architecture mask (archmask for short).
Our model has two stages: 1) a semantic layout diffusion model (\cref{fig:semdiff}a) for predicting the layout,  and 2) an attribute prediction model (\cref{fig:semdiff}b).
The two stages are trained separately and combined at the inference stage.

\subsection{Semantic Layout Generation}
\label{sec:method-semlayoutgen}

The first stage is based on the multinomial diffusion model~\cite{hoogeboom2021argmax}, shown in \Cref{fig:semdiff} $(a)$.
The multinomial diffusion model is a categorical discrete diffusion model designed for categorical data perfectly suited for our data representation.
We modify the multinomial diffusion model to make it a unified model conditioned on possible room masks for all room types.

During training, the input is a 2D segmentation map $\mathcal{S} \in \mathbb{R}^{H \times W \times K}$, where pixels indicate semantic ID. Each pixel value is a one-hot vector $\mathbf{x} \in \{0,1\}^K$, where $K$ is the number of categories.
We set $K=38$ semantic categories with 34 for object types, 3 for architecture elements (\semtype{floor}, \semtype{door}, \semtype{window}), and 1 for \semtype{void}. 
For conditioning, we specify the room type  $c_\text{room}$ and the room mask $\mathcal{A} \in \mathbb{R}^{H \times W}$ where $\mathcal{A}_{i,j} \in \{0,1,2,3\}$ denotes \semtype{void}, \semtype{floor}, \semtype{door}, and \semtype{window} respectively.
Note that we can also condition with the floor (i.e., the room mask without doors or windows), which can be expressed as a binary mask, or do unconditional generation of both the room architecture and the objects. 

We denote the noise pixel value at time step $t$ as $\mathbf{x}_t \in \{0,1\}^K$.
If $\mathbf{x_t}$ belongs to category $k$, then $\mathbf{x}_{tk} = 1$ and $\mathbf{x} _{tj}= 0$ for $j \neq k$.
The probability of $\mathbf{x_t}$ given $\mathbf{x_0}$ is 
\begin{equation*}
    q(\mathbf{x_t}|\mathbf{x_0}) = \mathcal{C}(\mathbf{x_t} | \Bar{\mathbf{\alpha}}_t\mathbf{x_0} + (1-\Bar{\mathbf{\alpha}}_t/K)
\end{equation*}
where $\mathcal{C}$ is the categorical distribution with parameters $\alpha_t = 1 - \beta_t$ and $\Bar{\alpha}_t = \prod_{\tau=1}^{t} \alpha_{\tau}$.

The objective of the multinomial diffusion model is to minimize the KL divergence of the categorical distribution between the generated data and ground truth, which is
\begin{equation*}
    L_{\text{MDM}} = \mathbb{E}_{\mathbf{x}, t} \left[ \text{KL}\left(q(\mathbf{x}_{t-1}|\mathbf{x_t}, \mathbf{x}_0), p(\mathbf{x}_{t-1}|\mathbf{x}_t, \mathcal{A}, c_\text{room})\right) \right]
\end{equation*}
where $q(\mathbf{x}_{t-1}|\mathbf{x_t}, \mathbf{x}_0)$ is the ground-truth categorical posterior and $p(\mathbf{x}_{t-1}|\mathbf{x}_t, \mathcal{A}, c_\text{room})$ is the predicted distribution at time $t-1$ given previous time $t$ distribution, room mask and room type condition. See \citet{hoogeboom2021argmax} for details of the multinomial diffusion model.

\Cref{fig:semdiff} shows how the room mask $\mathcal{A}$ is passed through an embedding layer and an MLP to get the room mask embedding, then added to the $\mathbf{x_t}$ embedding to control denoising so the generated layout can respect the input mask.
Furthermore, the room type $c_\text{room}$ embedding is added to the timestep embedding to allow generating the desired room type.
The embedding size for both the room mask and room type is 64.
These two conditions allow a unified model for all room types that generates the room semantic layout with different room types and masks as input.
By adding another control for the type of room mask used for conditioning (e.g., none, floor, arch), we can train a single unified model for all room and mask types (see \cref{sec:supp-mix}).

\subsection{Attribute Prediction}
\label{sec:method-apm}

The generated semantic layout only gives potential object instances and their 2D location on the projected plane.
Thus, we design an attribute prediction model (APM) to predict attributes ($s, p,  r$) for each object as shown in \Cref{fig:semdiff} (b).
Based on each instance mask, we obtain the 2D position and 2D size (width and length).
The APM network then predicts the vertical size and position.

The input of the APM is the semantic map and the extracted instance masks, while the output is the instance-level object attributes, including vertical size $s_{y_i} \in \mathbb{R}^1$, vertical position $p_{y_i} \in \mathbb{R}^1$, and orientation $r_i \in \{0,1,2,3\}$. 
We pass the semantic layout map and instance mask to the encoder $E_\text{layout}$ and $E_\text{mask}$ respectively to get the layout feature $f_\text{layout} \in \mathbb{R}^{128\times32\times32}$ and the instance mask feature $f_\text{mask} \in \mathbb{R}^{128\times32\times32}$.
We treat $f_\text{layout}$ as the query and $f_\text{mask}$ as the key and value to perform cross-attention to obtain the final object instance feature $f_\text{inst}\in \mathbb{R}^{128\times32\times32}$.
Finally, we use different prediction heads, consisting of a shared 2-layer MLP followed by a 1-layer MLP for each attribute, to predict attributes using $f_\text{inst}$.
During training, we use the ground-truth instance masks.
We use MSE loss for size and position heads, defined as \(L_s, L_p\), and cross-entropy loss for orientation loss \(L_r\). The total loss for the APM is a sum of the three losses: \(L_\text{APM} = L_s + L_p+ L_r\).

\subsection{Inference}
\label{sec:method-inference}

At inference, we follow the pipeline in \cref{fig:semdiff} to generate the final scene (see \cref{sec:supp-model-details} for details). 
\mypara{Semantic layout sampling.} 
Given the room mask $\mathcal{A}$ and room type $c_\text{room}$ as conditions, we sample a semantic layout map $\mathcal{S}$ using the semantic layout diffusion model. 
\mypara{Instance extraction.} Based on the generated semantic map $\mathcal{S}$, we extract instances using connected component analysis.
As there may be noisy pixels, we define category-specific size thresholds to filter out object instances that are too small.
We determine the category-specific thresholds by calculating the minimum ratio of object pixels to total room pixels per object type. 
If an object's pixel ratio falls below the type-specific threshold, it is deemed invalid and excluded from subsequent attribute prediction and object retrieval.
\mypara{Attribute prediction.} After we have both the semantic map and instance masks, we pass them to the attribute prediction model to predict attributes for each instance.
\mypara{Room construction and object retrieval.}
Lastly, we use the attributes to retrieve and place the objects. 
For unconditioned generation, we also construct the room based on the generated architecture mask.

\section{Experiments}
\label{sec:experiments}

We compare our \semdiff with two recent diffusion methods (\diffuscene, \midiff). 
See \cref{sec:supp-expr-details} for additional experimental and training details.

\mypara{Experimental protocol.} We use the experimental setup from prior work~\cite{paschalidou2021atiss,tang2023diffuscene}, but we revisit the evaluation protocol and show that rendering choices can greatly influence common evaluation metrics for scene generation (\cref{sec:supp-prelim-experiments}).  
We advocate for a specific set of choices (rendering using a unified color palette that groups semantically close objects together, with a floormask, and shows more object details).
\mypara{Dataset.} 
We train our model using the 3D-Front \cite{fu20213d} training split from ATISS \cite{paschalidou2021atiss} with 4616 rooms containing bedrooms, living rooms, and dining rooms.
We use 38 semantic classes (34 object types, 3 arch-element types, and \texttt{void}).

\begin{figure}
\includegraphics[width=\columnwidth]{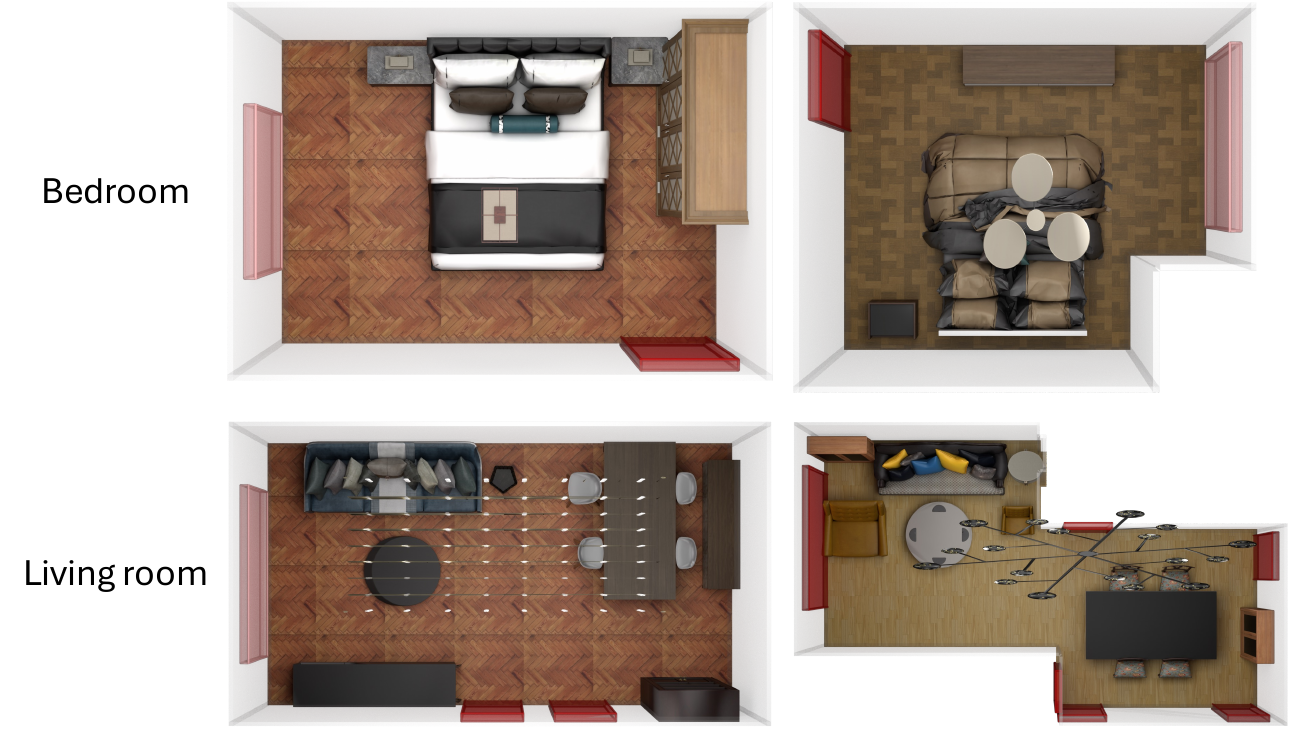}
\vspace{-18pt}
\caption{Example textured synthesized scenes with unconditional generation. 
We generate the room architecture together with the placement of the objects.
}
\vspace{-1em}
\label{fig:vis-texture}
\end{figure}

\mypara{Baselines.} 
We select two recent diffusion-based indoor scene synthesis methods~\cite{tang2023diffuscene,hu2024mixed} with training code for comparison.
We do not compare against CHoRD~\cite{su2025chord} as there is no code available, and report results for a pretrained PhyScene~\cite{yang2024physcene} model in \cref{sec:supp-physcene}.
Note that \midiff was originally designed to support floormask conditioning, while \diffuscene do not.
In addition, both works trained separate models for each room type.
For fair comparison, we add room‑type conditioning to prior methods (see \cref{sec:supp-adapt-prior-work} for implementation details), 
and consider three generation modes for room mask conditioning: no room mask (\emph{none}), conditioned on the \emph{floor}, and \emph{architecture}.

\begin{figure*}
\includegraphics[trim={0 0 0 5px},clip,width=\textwidth]{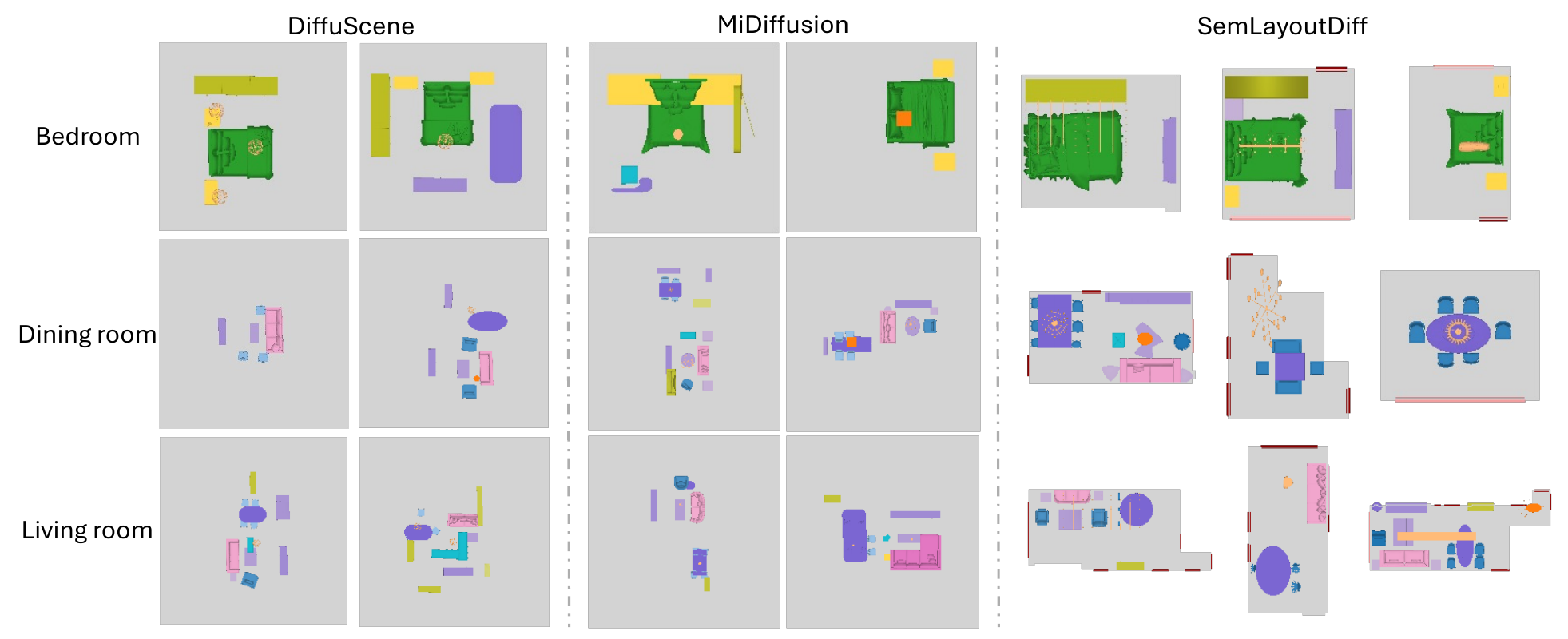}
\vspace{-20pt}
\caption{Comparison of scenes generated by prior methods and \semdiff \textit{without room mask (i.e., no floor or arch) conditioning}. Since prior methods (left) \textit{cannot} generate architectural elements, a square floor is used by default. Our method (right) generates feasible furniture placements aligned with its generated architectural layouts, leaving doors and windows unobstructed.
}
\vspace{-8pt}
\label{fig:vis-qua-uncon}
\end{figure*}

\begin{figure*}[ht]
\includegraphics[trim={0 5px 0 5px},clip,width=\textwidth]{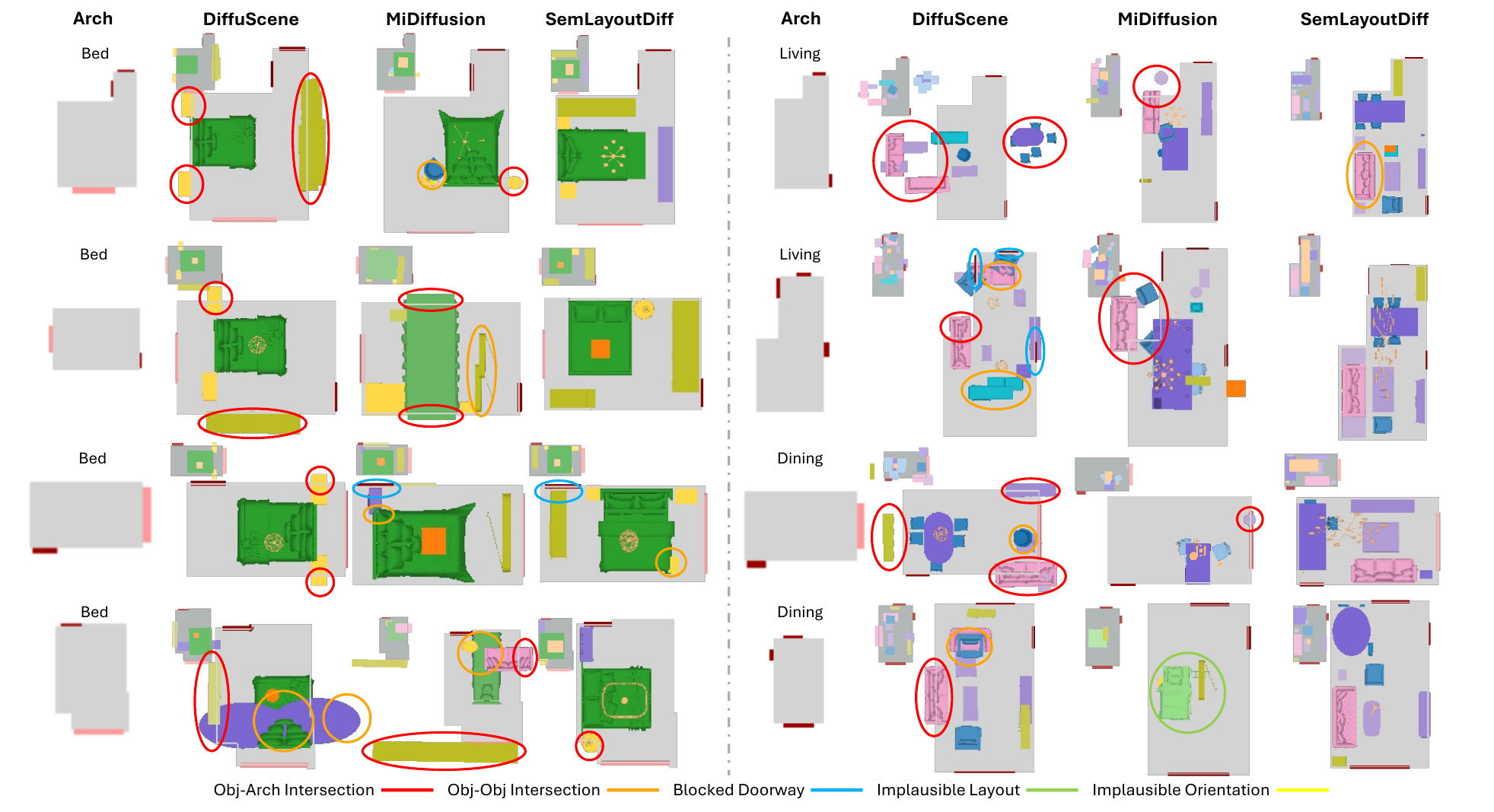}
\vspace{-20pt}
\caption{Comparison of generated scenes using different methods with \textit{arch mask conditioning}. 
For each scene, we show the bounding box layout before object retrieval, inset on the upper left.
Errors are indicated with color-coded circles (see \cref{sec:expr-user-study} for error types).
Our \semdiff generates scenes with fewer errors and respects architectural constraints by keeping furniture within room boundaries and maintaining clear spaces around doors (\red{red}) and windows (\pink{pink}), whereas \diffuscene and \midiff do not.
}
\vspace{-16pt}
\label{fig:vis-qua}
\end{figure*}

\subsection{Examples of generated scenes}
\label{sec:expr-examples}

To demonstrate our model's ability to generate plausible room layouts, we show examples of generated rooms  (objects and the room itself) in \cref{fig:vis-texture}. 
We also compare generated scenes from our \semdiff to those generated by \diffuscene and \midiff with the three generation modes: no room mask (\cref{fig:vis-qua-uncon}), conditioned on floor  (appendix \cref{fig:vis-qua-floor}), and on architecture (\cref{fig:vis-qua}). 
For the comparison, we use semantically colored renderings with orthographic projection to clearly show the layout 
(see \cref{sec:supp-prelim-experiments}).
These are the same renderings used in the quantitative evaluation and user studies.  
We provide additional qualitative results and renderings in \cref{sec:supp-expr-qualitative-results}.

\mypara{\semdiff can generate the architecture mask without conditioning.}
In \cref{fig:vis-qua-uncon}, we show examples of different rooms generated by our \semdiff and other methods.  
As prior methods (\diffuscene, \midiff) cannot generate the architecture mask, we show the scenes with a square floor.
In contrast, our proposed \semdiff effectively generates diverse and realistic arch masks (including floors, doors, and windows) even without explicit architectural conditioning. 
The scenes produced by \semdiff exhibit more plausible and varied room shapes and furniture placements, highlighting its capability to create coherent indoor scenes.

\mypara{\semdiff respects the architecture mask better.} From \cref{fig:vis-qua}, we observe that \diffuscene and \midiff struggle to place furniture with respect to architectural elements, often resulting in objects blocking doors, windows, or extending beyond room boundaries. In contrast, our proposed \semdiff consistently generates coherent, organized, and realistic furniture arrangements, effectively respecting architectural constraints such as doors, windows, and floor boundaries.

\begin{table}[t]
\centering
\caption{
\textbf{Distribution match to ground-truth scenes.}
We report the \fid, \kid, \sca, and \ckl with different levels of architecture conditioning: unconditioned (None), conditioned on the floor (Floor), and the architecture map (Arch). 
\textbf{Bold} indicates best results.
Our \semdiff outperforms other methods.
}
\vspace{-8pt}
\resizebox{\linewidth}{!}
{
\begin{tabular}{@{} l c rrrr @{}}
\toprule
Condition & Method &  \fid & \kid & \sca & \ckl \\
\midrule
\multirow{4}{*}{\shortstack[l]{None}} 
& \diffuscene \cite{tang2023diffuscene} & {125.46} & {88.09} & {99.75} & 24.70 \\
& \midiff \cite{hu2024mixed} & 100.54 & 41.27 & 97.35 & 28.31 \\
& \semdiff (Ours) & \textbf{93.93} & \textbf{10.72} & \textbf{96.76} & \textbf{17.21}\\
\midrule
\multirow{4}{*}{\shortstack[l]{Floor}} 
& \diffuscene \cite{tang2023diffuscene} & {90.82} & {30.13} & {94.43} & 23.93  \\
& \midiff \cite{hu2024mixed} & 91.79 & 26.23 & 97.27 & {23.64} \\
& \semdiff (Ours) & \textbf{81.79} & \textbf{14.52} & \textbf{89.83} & \textbf{5.99} \\
\midrule
\multirow{3}{*}{\shortstack[l]{Arch}} 
& \diffuscene \cite{tang2023diffuscene} & 88.47 & 30.22 & 95.02 & 20.42 \\
& \midiff \cite{hu2024mixed} & 93.51 & 31.06 & 95.28 & 36.97 \\
& \semdiff (Ours) & \textbf{71.06} & \textbf{8.65} & \textbf{86.96} & \textbf{4.78} \\ 
\bottomrule
\end{tabular}
}
\label{tab:quant-distmatch}
\end{table}
\begin{table}
\centering
\vspace{-6pt}
\caption{
\textbf{Physical plausibility of generated scenes.} We compare out-of-bounds (OOB) ratios at the scene (\oobscene) and object level (\oobobj), as well as the collision rate (\col), and navigability (\nav).
We do not include \diffuscene and \midiff for the \texttt{none} condition as they do not generate any room architecture, so most of the metrics are irrelevant.
}
\vspace{-8pt}
\resizebox{\linewidth}{!}
{
\begin{tabular}{@{} l c rrrr @{}}
\toprule

Condition & Method & 
\oobscene & \oobobj & \col & \nav \\
\midrule
None & \semdiff (Ours) &
2.04 & 0.46 & 17.70 & 95.30 \\

\midrule
\multirow{3}{*}{\shortstack[l]{Floor}} &
\diffuscene \cite{tang2023diffuscene} & 
69.93 & 28.92 & 37.68 & 94.99\\
& \midiff \cite{hu2024mixed} & 
70.17 & 31.83 & 35.92 & 96.02\\
& \semdiff (Ours) & 
\textbf{24.60} & \textbf{6.93} & \textbf{19.61} & \textbf{96.69}\\

\midrule
\multirow{3}{*}{\shortstack[l]{Arch}} &
\diffuscene \cite{tang2023diffuscene} & 
66.47 & 20.03 & 40.97 & 94.34 \\
& \midiff \cite{hu2024mixed} & 
60.68	& 34.66	&  51.62 & 96.19 \\ 
& \semdiff (Ours) & 
\textbf{16.63} & \textbf{6.62} & \textbf{16.13} & \textbf{96.36} \\ 
\bottomrule
\end{tabular}
\vspace{-1em}
}
\label{tab:quant-plausibility}
\end{table}

\begin{table*}[t]
\centering
\caption{
Comparison of different training strategies for \semdiff: single condition (per-masktype) vs mixed conditions.
}
\vspace{-8pt}
\small
{
\begin{tabular}{@{} l c rrrr rrrr @{}}
\toprule
 & & \multicolumn{4}{c}{Distribution} & \multicolumn{4}{c}{Plausibility}  \\
\cmidrule(l{0pt}r{2pt}){3-6} 
\cmidrule(l{0pt}r{2pt}){7-10} 
Condition & Training & \fid & \kid & \sca & \ckl & \oobscene & \oobobj & \col & \nav \\
\midrule
\multirow{2}{*}{\shortstack[l]{None}} 
& \cellcolor{lightgray}per-masktype & 
93.93 & \best{10.72} & 96.76 & 17.21 &
2.04 & 0.46 & 17.70 & 95.30 \\
& mixed & 
\best{72.51} & 11.27 & \best{92.81} & \best{5.37} &
\textbf{0.70} & \textbf{0.18} & \textbf{14.32} & \textbf{96.84}  \\
\midrule
\multirow{2}{*}{\shortstack[l]{Floor}} 
& \cellcolor{lightgray}per-masktype &
81.79 & 14.52 & 89.83 & 5.99 &
24.60 & 6.93 & 19.61 & \textbf{96.69} \\
& mixed & 
\best{76.24} & \best{10.60} & \best{86.04} & \best{4.38} &
\textbf{21.50} & \textbf{5.07} & \textbf{14.67} & 96.63 \\
\midrule
\multirow{3}{*}{\shortstack[l]{Arch}} 
& per-roomtype &
107.99 & 42.99 & 96.08 & 16.57 &
26.97 & 8.89 & 23.11 & 95.11 \\
& \cellcolor{lightgray}per-masktype &
\best{71.06} & \best{8.65} & \best{86.96} & 4.78 &
\textbf{16.63} & 6.62 & 16.13 & 96.36  \\
& mixed &
73.61 & 9.85 & 92.19 & \best{4.27} &
21.53 & \textbf{4.98} & \textbf{14.42} & \textbf{96.91}  \\
\bottomrule
\end{tabular}
}
\vspace{-8pt}
\label{tab:quant-training}
\end{table*}

\begin{table}
\centering
\caption{
User rankings for the three methods on 40 samples.}
\vspace{-8pt}
\small
{
\begin{tabular}{@{} c r r r r @{}}
\toprule
& & \multicolumn{3}{c}{Rank Distribution}\\
\cmidrule(l{0pt}r{2pt}){3-5} 
Method & AR$\downarrow$ & $1\text{st}\%$ & $2\text{nd}\%$ & 3\text{rd}$\%$ \\
\midrule
\diffuscene \cite{tang2023diffuscene} & 2.21  & 11.11 & 57.22 & 31.67  \\
\midiff \cite{hu2024mixed} & 2.55 & 8.47 & 28.33 & 63.19 \\
\semdiff (Ours) &  \best{1.25} & \best{80.42} & \best{14.44} & \best{5.14}\\
\bottomrule
\end{tabular}
\vspace{-1em}
}
\label{tab:user-study}
\end{table}
\begin{table}
\centering
\caption{
Fine-grained user evaluation.  Users were asked to assess errors found in generated scenes.}
\vspace{-4pt}
\resizebox{0.95\linewidth}{!}
{
\begin{tabular}{@{} c r r r r r @{}}
\toprule
 & \multicolumn{2}{c}{Intersection} & Blocked & \multicolumn{2}{c}{Implausible} \\
Method & Arch$\downarrow$ & Obj$\downarrow$ & Door$\downarrow$ & Layout$\downarrow$ & Orientation$\downarrow$ \\
\midrule
\diffuscene \cite{tang2023diffuscene} & 89.38 & 38.12 & 41.25 & 45.00 & 25.62 \\
\midiff \cite{hu2024mixed} & 80.62 & 60.00 & 27.50 & 78.12 & 55.62 \\
\semdiff (Ours) & \textbf{20.00} & \textbf{25.00} & \textbf{22.50} & \textbf{11.88} & \textbf{24.38} \\
\bottomrule
\end{tabular}
\vspace{-1em}
}
\vspace{-8pt}
\label{tab:user-study-detail}
\end{table}

\subsection{Quantitative evaluation}
\label{sec:expr-quantitative}

We evaluate the generated scenes by comparing against the distribution of real scenes, and evaluating the overall plausibility based on object collisions and how well the room mask is respected.
For each condition, we generate 1000 scenes for evaluation.

\mypara{Metrics.} 
For comparison of synthesized and real scenes, we follow ATISS~\cite{paschalidou2021atiss} and report the KL divergence (CKL) between object category distributions, and Fréchet inception distance (FID) \cite{heusel2017gans} and  Kernel inception distance (KID) \cite{binkowski2018demystifying} against the test set.  Following prior work, we report the CKL as CKL $\times 10^2$ and the KID as KID $\times 10^3$.  We also report the scene classification accuracy (SCA), which scores how well a trained classifier can distinguish synthesized scenes from real-world scenes. 

For SCA, the closer the score is to 50\%, the more difficult it is for the classifier to distinguish between generated and real scenes. We find that in our experiments, the selection of objects is severely limited, causing the SCA to easily identify generated scenes (see \cref{sec:supp-sca-analysis}).  
For the view-based metrics (KID, FID, SCA), we use a top-down semantic-colored rendering that clearly shows the objects and includes the architectural elements (see \cref{sec:supp-expr-reproduce-prior-work}).

To assess scene plausibility, we report the Out-of-Bounds (OOB) ratio~\cite{feng2023layoutgpt} (ratio of objects placed outside room boundaries).
We report both the percentage of scenes ($\text{OOB}_{S}$) and the percentage of objects ($\text{OOB}_{O}$) with OOB issues.
Finally, we report object collision rate (COL) and scene navigability (NAV) following SceneEval \cite{tam2025sceneeval}.

We report the performance of the different methods in \cref{tab:quant-distmatch,tab:quant-plausibility}, where we see that our \semdiff consistently outperforms other unified methods.

\mypara{\semdiff has a better distribution match to the ground-truth scenes.}
Compared to prior methods, \semdiff maintains closer visual and object distribution (see lower FID, KID, and CKL in \cref{tab:quant-distmatch}). 
The other methods have unrealistic object distributions, such as placing beds in living rooms or dining tables in bedrooms (e.g., in \cref{fig:vis-qua}, the \midiff-generated dining room at the bottom right has a bed in the middle; the \diffuscene-generated bedroom at the bottom left has two dining tables).
This issue arises because these models are trained on a mixed object distribution across room types, which can confuse learning. 
In contrast, our model treats scene layout generation as a semantic image synthesis problem, conditioned specifically by room type, thus facilitating more effective learning of distinct object distributions.

\mypara{Conditioning on floor and arch mask improves performance.}
\Cref{tab:quant-distmatch} shows that adding conditioning on the floor or arch mask significantly improves scene synthesis. 
The overall FID for \semdiff drops from 93.93 for no room mask to 81.79 with floor conditioning and 71.06 for arch conditioning.
The drop in FID, KID, and CKL shows that architecture information enables more realistic layouts matching the ground-truth distribution.

\mypara{\semdiff better respects architectural constraints and generates more plausible layouts.}
\Cref{tab:quant-plausibility} shows that our \semdiff consistently achieves significantly lower scene-level (OOB\textsubscript{S}) and object-level (OOB\textsubscript{O}) out-of-boundary ratios compared to prior methods, across all room types. For instance, in bedrooms, \semdiff reduces the OOB\textsubscript{S} from approximately $55\%$ (\diffuscene) and $60\%$ (\midiff) to $13.8\%$, and the OOB\textsubscript{O} from over $20\%$ to only $4\%$. These results demonstrate our method's superior ability to place furniture objects accurately within architectural boundaries.

\semdiff also consistently has the lowest object‑object collisions and maintains high navigability scores. 
For navigability, the only slight drop appears in bedrooms, where our navigability metric treats lights positioned below 2 meters as obstacles; such fixtures are common in practice, so this penalty does not reflect a real limitation of the layout. Overall, the results show that our method generates navigable layouts with less collisions.

\mypara{Mixed-condition training.}
In our experiments so far, we have trained a separate model for each room mask type (none, floor, arch).  We can also train one single unified model that can handle conditioning both on room type as well as different room mask types by training a single mixed condition model (see \cref{sec:supp-mix} for details). 

In \cref{tab:quant-training}, we show that with the mixed training, we generate scenes with slightly better distribution match and plausibility when conditioned on no room mask or just the floor mask.  When conditioning on arch mask, the per-masktype model has slightly better distribution match.  We also compare to training of a separate per-roomtype model.  Notably, we find that per-roomtype training results in generated scenes that are further from the ground-truth distribution (high FID / KID / CKL) and poor plausibility (more out-of-bounds and collisions).  The per-roomtype training suffers from both poor performance as well as the need for multiple models.  
By including all room types and mixing the conditioning allows the models to be trained on more data and learn the overall distribution better.    
To handle the different conditioning types, the per-roomtype training (as common in prior work) would need $M 
\times R$ separate models where $R$ is the number of room-types and $M$ is the number of mask types, vs $M$ models for the per-masktype training, and $1$ model for our proposed mixed condition training.

\subsection{User study}
\label{sec:expr-user-study}

\begin{figure}
\includegraphics[width=\linewidth]{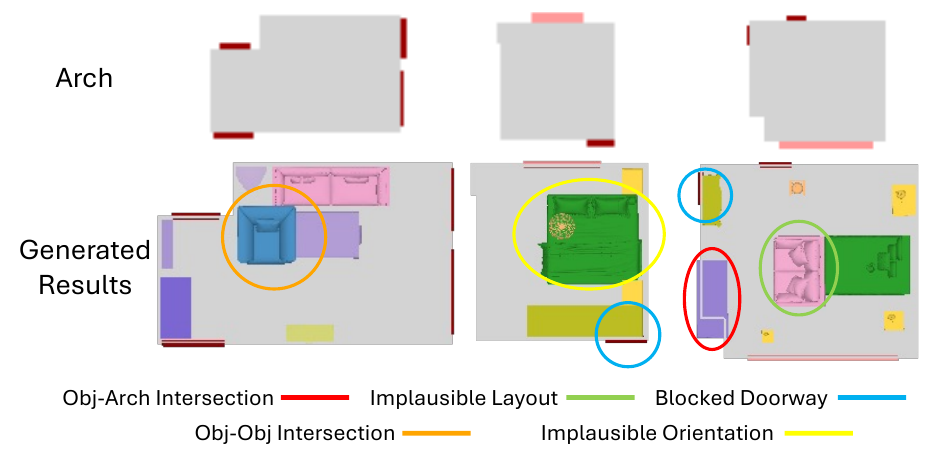}
\vspace{-20pt}
\caption{
Examples of error cases for user study from \semdiff-generated scenes with arch mask conditioning.
}
\vspace{-8pt}
\label{fig:vis-error-types}
\end{figure}

To further compare the quality of the generated results, we conducted two user studies. 
As prior methods cannot generate the architecture, for fairer comparison we focus on generation conditioned on the arch mask.
The first study asked participants to rank three generated scenes conditioned on the same architectural mask with different models according to plan adherence and overall layout plausibility.  In \cref{tab:user-study}, we report average ranking (AR) and the frequency of 1st, 2nd, and 3rd places. The second study examined five specific error types: object‑architecture intersection or out-of-bounds, object‑object intersection, blocked doorway, implausible layout, and implausible orientation (\cref{fig:vis-error-types}). We report the percentage of scenes in which each error appears (\cref{tab:user-study-detail}).
See \cref{sec:supp-user-study-details} for more details on the user studies.

\mypara{\semdiff has the best performance.}
As shown in \cref{tab:user-study},
\semdiff achieved the best average ranking (AR=$1.25$), with users selecting it as the best  $80\%$ of the time,  and only $~5\%$ ranking it last.
In the fine-grained evaluation (\cref{tab:user-study-detail}), \semdiff also outperform \diffuscene and \midiff with users reporting lower issue rates in all 5 cases.
Both user studies reinforces the findings from the automatic metrics: \textit{\semdiff generates the most plausible scenes as judged by people}.

\subsection{Discussion}
\label{sec:discussion}

\begin{figure}[t]
\includegraphics[width=\columnwidth]{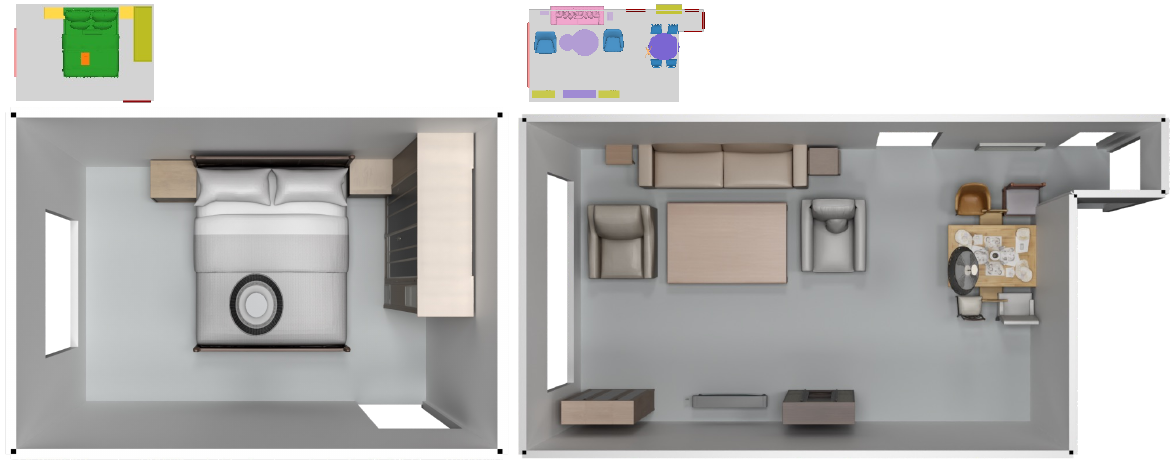}
\vspace{-20pt}
\caption{Example of our synthesized scenes with TRELLIS-generated objects. Top left is the top-down rendering with 3D-FUTURE retrieved objects using our STK rendering, and bottom is the TRELLIS-based rendering with Blender.
}
\vspace{-1em}
\label{fig:vis-trellis}
\end{figure}

\mypara{Scene synthesis with generated object.}
In \cref{fig:vis-trellis}, we show that layouts from \semdiff can be converted into 3-D scenes by pairing them with a 3D object generation model. 
Specifically, we use TRELLIS \cite{xiang2024structured}, prompting it with ``A \verb|<category>|'' to create a 3D object for each category in the layout. 
Then each mesh is scaled, translated, and rotated according to the predicted attributes. 

\mypara{Limitations.} Despite the strong performance of our \semdiff in unified indoor scene synthesis, it has several limitations. 
First, our method does not leverage object shape or contextual information, which could improve attribute predictions, such as orientation, and enhance object consistency.
Second, \semdiff currently does not support conditioning on text or a partial scene, limiting the level of user control. 
Third, relying on the small-scale 3D-FRONT dataset, which includes fewer than 5000 valid training rooms, constrains our model's potential; incorporating more diverse and extensive 2D and 3D scene data could enable the generation of more complex, multi-room layouts. 
Lastly, since our method uses top-down semantic maps, it struggles with vertical occlusion and hierarchical object arrangements. Future work can address this using a full 3D semantic voxel grid~\cite{yang2021indoor} or integrating semantic maps from multiple viewpoints, enabling better vertical layout generation.

\section{Conclusion}
\label{sec:conclusion}

In this work, we represent the indoor scene as the 2D semantic layout map with instance attributes to better capture the spatial relationship between objects and generalize to different room types.
With this representation, we propose a novel indoor scene synthesis model called \semdiff by leveraging the discrete denoising diffusion probabilistic model.
Unlike most of the previous methods that are designed to train the models separately for different room types, our model is a unified model that can be generalized to different room types.
Our model demonstrates flexibility in generation modes: it can operate without any room mask, with floor mask conditioning, or with full architectural mask conditioning. 
The ability to incorporate architectural elements (doors and windows) beyond just the floor mask is particularly valuable for indoor scene synthesis, as these elements impose important constraints on furniture placement and overall room layout.
Compared to prior works, our approach exceeds the performance in unified room-type synthesis and can better place the objects in the conditioned room mask.
Our proposed scene representation and corresponding method are important to the future of 3D generative modeling.
We hope our work can inspire the following work to generate more complex and realistic scenes.

\section*{Acknowledgments}
\label{sec:ack}

This work was funded in part by a CIFAR AI Chair and NSERC Discovery Grants, and enabled by support from the \href{https://alliancecan.ca/}{Digital Research Alliance of Canada} and a CFI/BCKDF JELF.
We thank Ivan Tam for help with running SceneEval; Yiming Zhang and Jiayi Liu for suggestions on figures; Derek Pun, Dongchen Yang, Xingguang Yan, and Manolis Savva for discussions, proofreading, and paper suggestions.  We also thank the anonymous reviewers for their feedback.

{
\small
\bibliographystyle{ieeenat_fullname}
\bibliography{main}
}

\appendix
\clearpage
\newpage
\maketitlesupplementary
\appendix
\setcounter{figure}{0}
\renewcommand{\thefigure}{\thesection.\arabic{figure}}

In this supplement, we present model details (\cref{sec:supp-model-details}), additional information about our experimental setup (\cref{sec:supp-expr-details}), preliminary experiments to investigate the impact of rendering changes to evaluation metrics (\cref{sec:supp-prelim-experiments}), and additional results (\cref{sec:supp-additional-results}).

\section{Model details}
\label{sec:supp-model-details}

\mypara{Instance extraction.}
Given the semantic map, we extract instances by finding connected components (with horizontal and vertical grid cells as neighbors) and identifying each connected region with the same category label as an individual object instance.
Specifically, we use the \textit{label} function in \textit{scipy} library to process the binary mask for each category.

\mypara{Object retrieval.}
For each extracted object instance, we retrieve a corresponding 3D model from a subset of 3D-FUTURE assets, which are used in the filtered rooms,
by computing the mean squared error (MSE) between the predicted 3D size of the instance and the sizes of available assets within the same category. The asset with the lowest MSE is selected as the best match:
\begin{equation*}
    \mathbf{a}^* = \arg\min_{\mathbf{a} \in \mathcal{A}_c} \; \mathrm{MSE}(\mathbf{s}_p, \mathbf{s}_a)
\end{equation*}
where
\begin{equation*}
    \mathrm{MSE}(\mathbf{s}_p, \mathbf{s}_a) = \frac{1}{3} \sum_{d \in \{x, y, z\}} (s_{p,d} - s_{a,d})^2,
\end{equation*}
and $\mathbf{s}_p$ and $\mathbf{s}_a$ denote the predicted and candidate asset sizes respectively, and $\mathcal{A}_c$ is the set of assets for category $c$.

\mypara{Room construction.}
Based on the generated semantic map, we extract the outline of the floor for two cases: (1) For floor- or architecture-conditioned generation, we directly use the floor or architecture mask, mapping each pixel to real-world coordinates on the xy-plane at a height of 0, with each pixel representing 0.01 meters. (2) For unconditional generation, we treat all non-zero pixels as floor, and similarly, extract the positions of doors and windows from their respective mask regions. 
We assume that doors are 2 meters tall and windows are rectangular, with a vertical offset of 0.5 meters to 2 meters.  
We construct the 3D geometry of the room by taking the room outline and creating planar walls, assuming a wall height of 3 meters, and cutting out rectangular holes for the doors and windows.

\section{Experimental details}
\label{sec:supp-expr-details}

We provide details on our experimental setup, including the data preprocessing and filtering (\cref{sec:supp-data-processing}) for our experiments, the unified set of object categories (\cref{sec:supp-unified-categories}), how baseline models are adapted and trained (\cref{sec:supp-adapt-prior-work}), training details for our \semdiff (\cref{sec:supp-training-details}), discussion of compute and memory cost (\cref{sec:supp-model-efficiency}), evaluation metric details (\cref{sec:supp-metric-details}), and details of our user study (\cref{sec:supp-user-study-details}).  

\subsection{Data processing}
\label{sec:supp-data-processing}
We preprocess the data and use BlenderProc \cite{denninger2023blenderproc2}, to render top-down views of the scenes. 
For the semantic scene layout, we render both instance and semantic masks.  We follow the steps below for processing the data: 
\begin{enumerate}
   \item We first extract rooms from scenes based on their room types (e.g. bedroom, living room, dining room) and  
    filter the data following DiffuScene \cite{tang2023diffuscene}. 
    Specifically, we remove rooms with unnatural dimensions (e.g., overly large sizes or extreme heights), discard infrequent objects appearing in fewer than 15 rooms, and exclude scenes containing overlapping or misplaced objects, as well as rooms with an insufficient or excessive number of objects. 
    \item  After this filtering process, the dataset includes 4041 bedrooms, 900 living rooms, and 813 dining rooms.
    For the unified dataset, we combine all room types and split the processed data into 70\% training, 10\% validation, and 20\% test following prior work~\cite{paschalidou2021atiss,tang2023diffuscene}. 
    \item We render the top-down 2D semantic layout map using orthographic projection, where the camera is positioned at the center top of the room.  To ensure that each pixel on the map represents a consistent physical unit $s$, we compute the desired image dimensions in pixels $(w_\text{img}, h_\text{img})$ from the input room width $w_\text{room}$ and length $l_\text{room}$ in meters.  As we render a top-down view, the image dimensions are given by:
    $
    (w_\text{img}, h_\text{img}) = (w_\text{room}, l_\text{room}) / s
    $.
    We set the Blender orthographic scale ($OS$)\footnote{The Blender orthographic scale specifies the larger dimension in scene units (meters)} to the maximum of the room width and length ($OS = \max(w_\text{room}, l_\text{room})$). 
    We pad the image to $1200\times1200$ (corresponding to $12\text{m}\times12\text{m}$) for later training and inference.
    \item We then extract the objects' category, size, vertical position (offset from floor), and orientation based on the annotations. The offset is determined by measuring the maximum distance from the object's bounding box to the floor. 
\end{enumerate}

\subsection{Unified object categories}
\label{sec:supp-unified-categories}

We use the same object categories following ATISS and DiffuScene. In those works, different sets of object categories were used for different room types.
To obtain a unified set of object categories, we combine object types from different room types into a single unified set. 
To ensure clarity regarding the object categories used in our analysis, we provide a list of the unified categories as follows: \semtype{kids bed, single bed, double bed, corner side table, round end table, coffee table, console table, tv stand, desk, dressing table, table, dining table, stool, dressing chair, dining chair, chinese chair, armchair, chair, lounge chair, loveseat sofa, lazy sofa, sofa, multi seat sofa, chaise longue sofa, l shaped sofa, nightstand, shelf, bookshelf, children cabinet, wine cabinet, cabinet, wardrobe, pendant lamp, ceiling lamp, floor, door, window, void}.

\subsection{Adapting prior methods}
\label{sec:supp-adapt-prior-work}

For fair comparison, we adapt \diffuscene and \midiff to train unified models that handle all room types, instead of their original separate models per room type.
We modify these methods to accept both architecture plan and room type as conditioning inputs. 
To allow conditioning by room type, we adapt the models so that the room type is passed in as a one-hot vector and then converted into a dense representation using a Multi-Layer Perceptron (MLP). 
We also modified \diffuscene and \midiff to incorporate architectural plan conditioning.
\textbf{\diffuscene} \cite{tang2023diffuscene} represents scenes as fully connected graphs, with each node corresponding to an object and its attributes.
It also uses dense representations for room type and floorplan as instance-level attributes, supplemented with positional encodings to position objects accurately. 

During inference, this ensures the generation of coherent scenes aligned with the provided floorplan and room type. Without floor conditioning, the floorplan vector is omitted during training.
To adapt this method, we represent the architectural plan $\mathcal{A}$ as a single-channel image and extract architecture features using a ResNet-18 encoder.

\textbf{\midiff} \cite{hu2024mixed} represents scenes as sequences of objects with attributes, similar to \atiss and \diffuscene. 
However, it treats object attributes as both discrete (object categories) and continuous (size, orientation, position). 
\midiff incorporates additional spatial information by representing the floor as a set of 2D points with normals, following the approach described in \cite{wei2023lego}. It employs a PointNet encoder, trained from scratch, to extract features from the floormask.
To incorporate architectural constraints, we extract points and normals from the architecture mask separately, including floor, door, and window information. 
We encode the door and window features using a PointNet encoder and sum these features together. 
The summed feature is then concatenated with the floormask feature obtained from the original \midiff approach to form the architecture conditioning feature for the model.

\subsection{\semdiff Training details}
\label{sec:supp-training-details}

\mypara{Semantic Layout Generation.} We train the diffusion model using the Adam optimizer~\cite{kingma2014adam} with 4000 diffusion steps and a batch size of 64. We use a constant learning rate of 1e-4.  Training was conducted on a single 48G A40 GPU over 1000 epochs, completed in 24 hours.

\mypara{Attribute Prediction Module.} We train the APM using the Adam optimizer~\cite{kingma2014adam} with a weight decay of 1e-3 and a learning rate of 1e-5. The model processed images at a resolution of $1200\times1200$ using a batch size of 24. Training was conducted on a single 48G A40 GPU for 300 epochs, taking $\sim$36 hours.

\begin{table}
\centering
\caption{
Compute and memory cost of different models 
}
\vspace{-8pt}
\resizebox{\linewidth}{!}
{
\begin{tabular}{@{} c rr rr @{}}
\toprule
 & \multicolumn{2}{c}{Train} & \multicolumn{2}{c}{Inference} \\
\cmidrule(l{0pt}r{2pt}){2-3} \cmidrule(l{0pt}r{2pt}){4-5}
 & Time & Mem & Time & Mem \\
\midrule
\diffuscene & 52h & 5G (batch=128)  & 25.0s & 0.5G\\
\midiff & 120h & 4G (batch=128) & 11.0s & 0.4G\\ 
\semdiff & 60h & 11G (batch=24) & 18.2s & 1.4G\\
\bottomrule
\end{tabular}
}
\label{tab:model-efficiency}
\end{table}

\subsection{Model efficiency}
\label{sec:supp-model-efficiency}
Under the architecture-mask conditioning and unified setting, \semdiff samples a room in $18.2$ seconds: $18$s for the $4,000$ diffusion steps and $0.2$s for attribute prediction. Its semantic layout map and attribute models are trained in $24$h and $36$h, respectively; trained in parallel, the wall-clock time is $36$h. \midiff needs $11$s per scene with a $1,000$-step schedule but $120$h for training. \diffuscene spends $25$s per scene with the same $1,000$ steps and trains in $52$h. Hence, \semdiff attains moderate inference latency and a lower training time than \midiff.
See \cref{tab:model-efficiency} for a summary. 

\subsection{Evaluation metric details}
\label{sec:supp-metric-details}

\begin{figure*}[t]
\centering
\includegraphics[trim={0 10px 0 0},clip,width=0.9\textwidth]{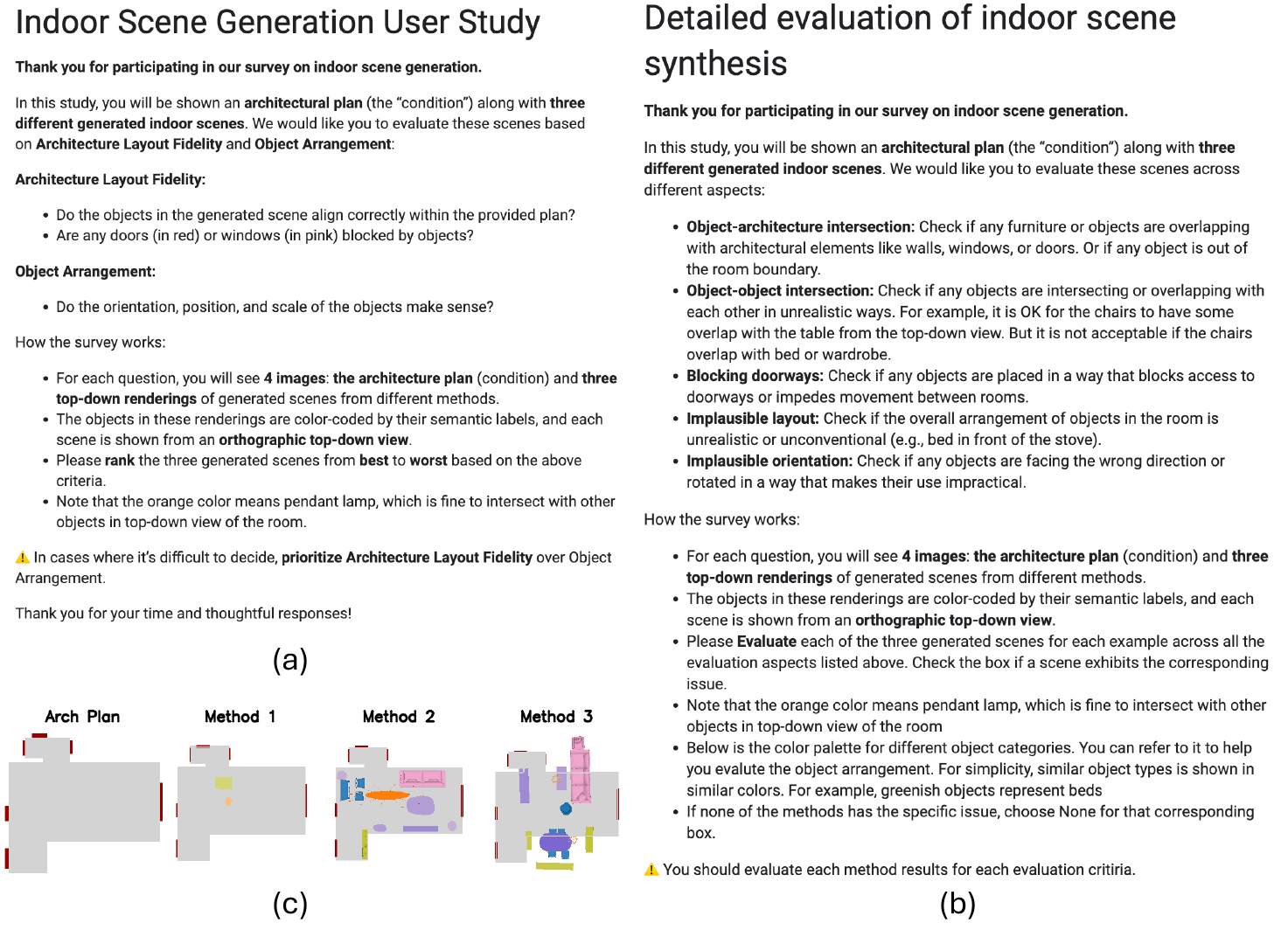}
\caption{
User study details. (a) shows the detailed instruction of our overall quality evaluation user study; (b) shows the detailed instruction of our fine-grained user study to evaluate specific error types in the generated rooms; (c) shows an example of the provided sample, including arch plan and randomly ordered generated rooms from three different methods.
}
\label{fig:vis-user-study}
\end{figure*}

To assess the quality of generated scenes, we use a set of quantitative metrics that cover visual fidelity, semantic consistency, category distribution, spatial plausibility, and navigability. 
Following prior work~\cite{paschalidou2021atiss,tang2023diffuscene}, we adopt FID, KID, and SCA for image-based evaluation and CKL for measuring the category distribution alignment. From SceneEval \cite{tam2025sceneeval}, we adopt COL to assess object-level collisions and NAV to evaluate scene navigability. In addition, we implement our own OOB-scene and OOB-object metrics to check whether objects extend beyond the floor boundary. Together, these metrics provide a complementary and thorough evaluation of scene generation quality. 

\mypara{FID and KID.}  
Fréchet Inception Distance (FID) and Kernel Inception Distance (KID) are standard image generation evaluation metrics for measuring how well the generated images match the distribution of the training images. 
For scene generation, rendered images are typically used.
FID computes the distance between Gaussian distributions fitted on features of generated and real images:
\begin{equation*}
\mathrm{FID} = \| \mu_r - \mu_g \|_2^2 + \mathrm{Tr}\!\left( \Sigma_r + \Sigma_g - 2(\Sigma_r \Sigma_g)^{1/2} \right),
\end{equation*}
where $(\mu_r, \Sigma_r)$ and $(\mu_g, \Sigma_g)$ are the mean and covariance of real and generated features.  
KID instead uses the squared Maximum Mean Discrepancy (MMD) with a polynomial kernel:
\begin{equation*}
\mathrm{KID} = \mathbb{E}[k(x_r,x_r')] + \mathbb{E}[k(x_g,x_g')] - 2\mathbb{E}[k(x_r,x_g)],
\end{equation*}
where $k$ is a polynomial kernel and expectations are taken over real ($x_r$) and generated ($x_g$) features.  

\mypara{SCA.}  Scene classifier accuracy (SCA) measures whether generated images preserve semantic consistency, by classifying each generated scene image into a room type and reporting the classification accuracy.
For the SCA classifier, the classifier is trained for each set of generated scenes.  
The classifier is trained to distinguish the ground truth training scenes from 500 different synthesized scenes (half of 1000).  The trained classifier is then evaluated on the other 500 synthesized scenes and the ground truth test scenes.

\mypara{CKL.}  
Category KL divergence (CKL) evaluates the alignment of category distributions between generated and real scenes. 
Let $p(c)$ and $q(c)$ be the normalized category frequencies (GT and generated). We compute:
\begin{equation*}
\mathrm{CKL}=\sum_{c} p(c)\,\log\!\frac{p(c)+\varepsilon}{q(c)+\varepsilon},\quad \varepsilon=10^{-6}.
\end{equation*}
A lower value indicates that the generated scenes reproduce the real-world category frequency more faithfully.

\begin{table*}
\centering
\caption{
Sanity checks for \diffuscene under the condition that uses \diffuscene semantic class-based rendering and w/o floor.  We compare our retrained model against that of the results reported in the original paper.}
\resizebox{\linewidth}{!}
{
\begin{tabular}{@{} c rrrr rrrr rrrr @{}}
\toprule
& \multicolumn{4}{c}{Bedroom} & \multicolumn{4}{c}{Dining room} & \multicolumn{4}{c}{Living room} \\
\cmidrule(l{0pt}r{2pt}){2-5} \cmidrule(l{0pt}r{2pt}){6-9} \cmidrule(l{0pt}r{2pt}){10-13}
Method & \fid & \kid & \sca & \ckl & \fid & \kid & \sca & \ckl & \fid & \kid & \sca & \ckl  \\
\midrule
\diffuscene (paper) & 17.21 & 0.7 & 53.52 & 0.35 &
32.60 & 0.72 & 55.50 & 0.22 & 
36.18 & 0.88 & 57.81 & 0.21 \\
\diffuscene (pretrained) & 26.66 & 2.77 & 63.03 & 0.39 & 30.85 & 4.94 & 60.32 & 0.28 & 27.95 & 3.30 & 54.89 &  0.31 \\
\diffuscene (retrained) & 28.18 & 0.19 & 53.33 & 0.41 &
41.49 & 0.60 & 50.07 & 0.27 &
43.37 & 0.46 & 53.88 & 0.44 \\
\bottomrule
\end{tabular}
}
\label{tab:diffuscene-baseline}
\end{table*}

\mypara{OOB-scene and OOB-object.}  
The out-of-boundary (OOB) metrics detect objects that extend beyond the floor boundary based on their 3D bounding boxes. An object is marked OOB if any part lies outside the floor polygon.  
OOB-object reports the ratio of such objects over all objects, while OOB-scene reports the ratio of scenes that contain at least one OOB object. These metrics capture violations of spatial plausibility.

\mypara{COL.}  
Collisions (COL) evaluates whether objects in a scene physically intersect with each other. 
We perform pairwise mesh-level intersection tests among all objects, and count the percentage of objects involved in any collision. 
A lower COL indicates that the generated layout respects basic physical plausibility.

\mypara{NAV.}  
Navigability (NAV) measures how easily an agent can move through the scene. 
We project the scene onto the floor plane, mask out the regions occupied by objects, and compute the largest connected free space. 
The ratio of this largest region to the total free space indicates navigability. 
A higher NAV means the layout allows smooth movement and avoids isolated or blocked areas.

\subsection{User study details}
\label{sec:supp-user-study-details}

We conduct two separate user studies: a) an overall assessment of scene quality and b) a more fine-grained evaluation of errors in the generated scenes. 
In \cref{fig:vis-user-study} (a,b), we show the full instructions we provided the users for the two studies.  For each study, we show the partipants generated scenes from the three methods (the baseline methods DiffuScene and MiDiffusion, and our SemLayoutDiff) with a random ordering of the methods, and the method names hidden to prevent bias.
We show the input architectural plan and the semantic  coloured, top‑down scenes generated by each of the three methods in random order (\cref{fig:vis-user-study} (c)).  We used the same $40$ arch + scenes for the two studies.
We recruited 18 participants to conduct the overall quality evaluation study and 4 participants for the second fine-grained evaluation study.

\mypara{Overall quality evaluation.} 
Participants ranked the scenes 1 (best) to 3 (worst) based on the overall quality. If the quality is hard to rank, the participants first judged how well each method's result fits in the architecture plan; then they compared overall object arrangement (\cref{fig:vis-user-study} (a)).

\mypara{Fine-grained evaluation.} 
For this study, participants used a grid of checkboxes to indicate what errors were present for each scene.  We asked users to assess whether the scene had any of the following five types of errors: object architecture overlap, object object overlap, blocked doorway, implausible layout, and wrong orientation.
Note that object architecture overlap also includes cases where an object is placed entirely outside the wall boundary.
Participants were asked to select every issue they detected for each scene or chose “none” if none applied (\cref{fig:vis-user-study} (b)).

\section{Preliminary experiments}
\begin{figure*}
\centering
\setkeys{Gin}{width=\linewidth}
\begin{tabularx}{\linewidth}{@{} Y | Y  Y  Y  Y  @{}}
\toprule
 & Bedroom & Bedroom & Dining room & Living room \\
 \midrule
\shortstack{simple3dviz\\(original palette,\\ overhead point-light)} & 
\imgclip{20}{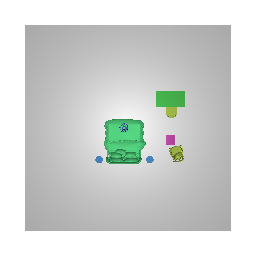} &
\imgclip{20}{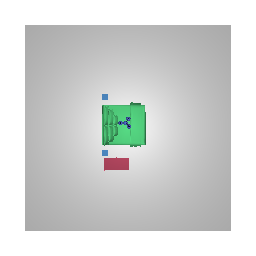} &
\imgclip{20}{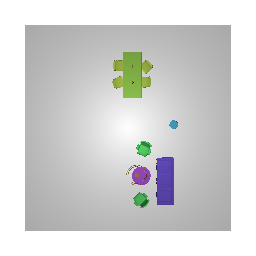} &
\imgclip{20}{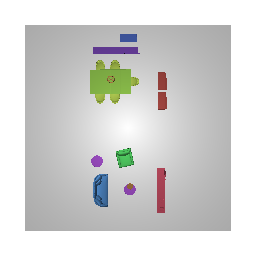}\\
\midrule
\shortstack{STK\\(new palette,\\constant illumination)} & 
 \imgclip{20}{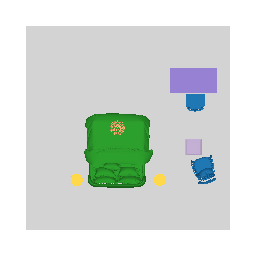} &
\imgclip{20}{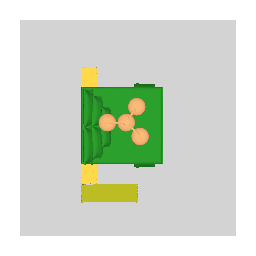} &
\imgclip{20}{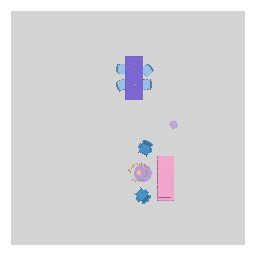} &
\imgclip{20}{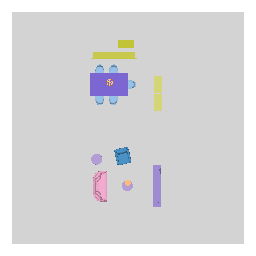}\\
\midrule
\shortstack{STK\\(with architecture)} &
 \imgclip{20}{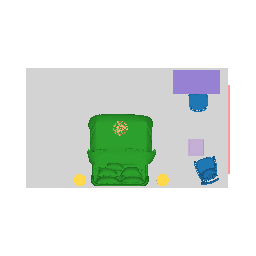} &
\imgclip{20}{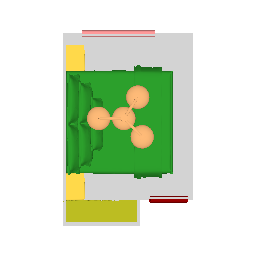} &
\imgclip{20}{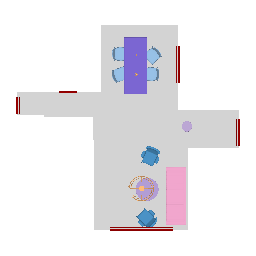} &
\imgclip{20}{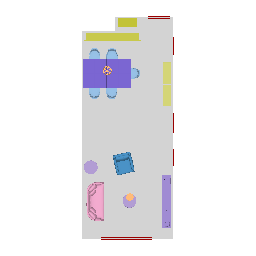}\\
\bottomrule
\end{tabularx}
\vspace{-8pt}
\caption{Examples of top-down renderings using simple3dviz with original palette (top), STK with updated palette (mid), STK with updated palette and actual architecture (bottom).  
In our new color palette (see \cref{fig:vis-color-palette}), we use similar colors for similar object categories (e.g. \texttt{single\_bed} vs \texttt{double\_bed}), resulting in more consistent coloring of objects across rooms (e.g. different types of sofa are colored similarly) with semantically different types of objects being more distinct (e.g. chairs around tables in column 3 are more sharply contrasted in the bottom row vs the top).
We also eliminate the point light, and visualize the actual architectural elements (floor, door, window).
These rendering changes aims to eliminate the lighting effect, and focus on the semantics of the objects and their placement with respect to the architecture elements.
}
\label{fig:vis-diff-color-qua}
\end{figure*}
\begin{figure*}
\centering
\setkeys{Gin}{width=\linewidth}
\begin{tabularx}{\linewidth}{@{}  Y  Y  Y  @{}}
 \imgclip{0}{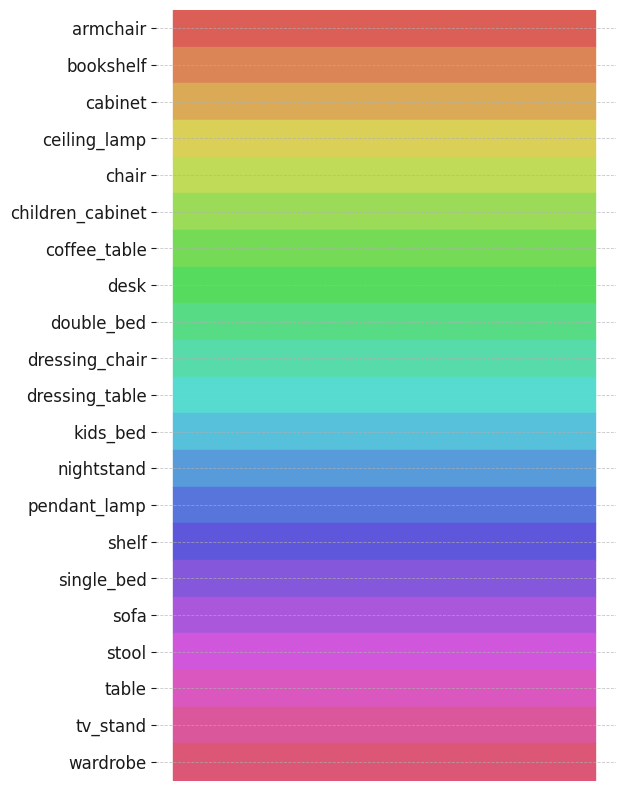} & \imgclip{0}{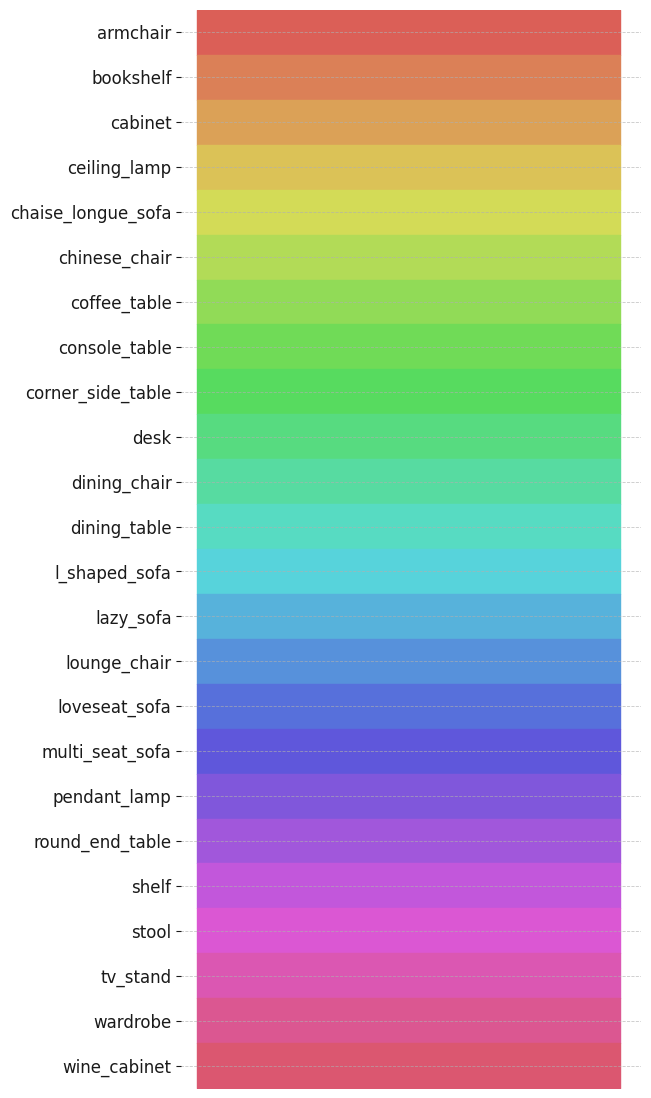} & \imgclip{0}{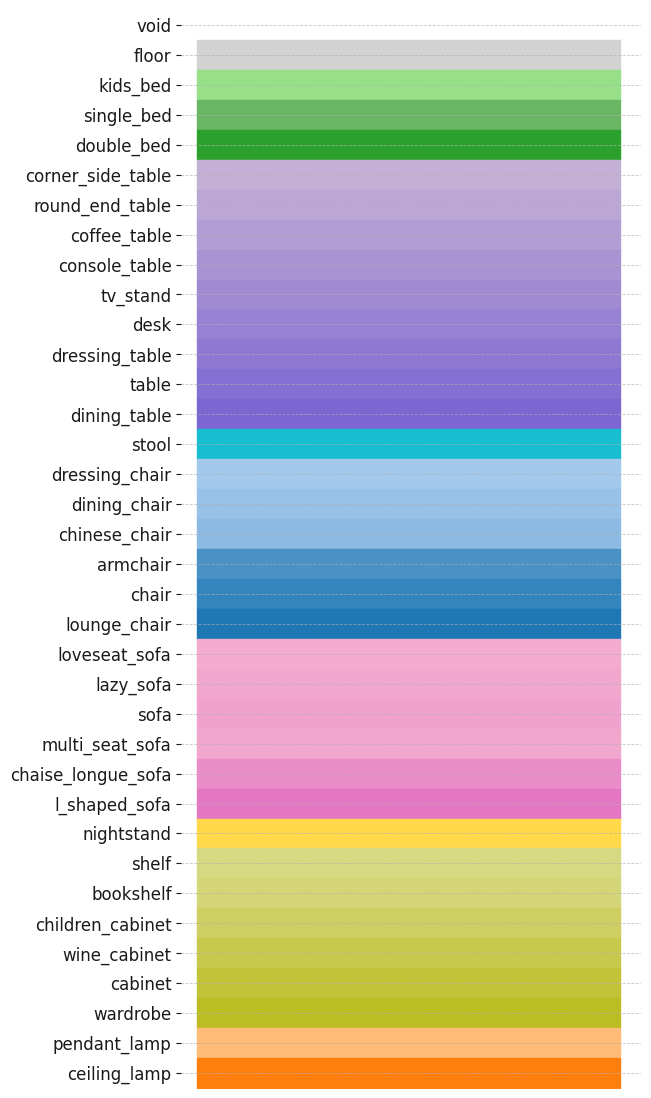}\\
 \diffuscene bedroom & \diffuscene living and dining room & \semdiff \\
\end{tabularx}
\vspace{-8pt}
\caption{The original color palette used by DiffuScene~\cite{tang2023diffuscene} vs the unified color palette that we use for rendering semantic colored scenes for evaluation.  Compared to the original color paletted, we tried to use similar colors for semantic classes that were more similar to each other.  For instance, we used shades of orange for both \texttt{pendant\_lamp} and \texttt{ceiling\_lamp} and shades of purple for various types of tables.  In addition, we tried to ensure classes that are semantically more different used more distinct colors (e.g. we used blue for chairs vs very close shades of aqua blue for both \texttt{dining\_chair} and \texttt{dining\_table}). Unlike DiffuScene, our color palette is also unified across room types (i.e. we use the same palette for all room types vs having separate colors palettes for bedroom and living/dining room).
}
\vspace{-8pt}
\label{fig:vis-color-palette}
\end{figure*}
\begin{table*}
\centering
\caption{
Comparison of how metrics change with different styles for DiffuScene (with simple-3dviz, unconditioned).
}
\vspace{-8pt}
{
\begin{tabular}{@{} lc rrr rrr rrr @{}}
\toprule
 & & \multicolumn{3}{c}{Bedroom} & \multicolumn{3}{c}{Dining room} & \multicolumn{3}{c}{Living room} \\
\cmidrule(l{0pt}r{2pt}){3-5} \cmidrule(l{0pt}r{2pt}){6-8} \cmidrule(l{0pt}r{2pt}){9-11}
Coloring (palette) & Floor & \fid & \kid & \sca & \fid & \kid & \sca & \fid & \kid & \sca  \\
\midrule
Semantic (DiffuScene) & none & 26.66 & 2.77 & 63.03 & 30.85 & 4.94 & 60.32 & 27.95 & 3.30 & 54.89 \\
Semantic (ours) & none  & 27.32 & 2.99 & 71.02 & 35.23 & 5.93 & 58.39 & 32.12 & 4.17 & 55.71 \\
Semantic (DiffuScene) & square & 32.98 & 0.49 & 54.16 & 45.79 & 0.86 & 49.37 & 45.37 & 0.81 & 51.67 \\
Semantic (ours) & square & 34.52 & 1.12 & 58.01 & 49.69 & 0.51 & 49.86 &  50.10 & 0.45 & 53.41 \\
Textured & square & 46.55 & 0.34 & 50.43 & 63.03 & 0.39 & 44.43 &  60.27 & 0.47 & 48.75 \\
\bottomrule
\end{tabular}
}
\label{tab:check-textured-vs-semantic-diffuscene}
\end{table*}
\begin{table*}
\centering
\caption{
Comparison of ATISS~\cite{paschalidou2021atiss}, DiffuScene~\cite{tang2023diffuscene}, and LayoutGPT~\cite{feng2023layoutgpt} using semantically rendered scenes generated with our colour palette against the colour palette used in DiffuScene, and using textured assets.  All scenes were generated using default settings from prior work and rendered with a square floor plan with simple-3dviz. 
}
\vspace{-8pt}
\resizebox{\linewidth}{!}
{
\begin{tabular}{@{} cc rrr rrr rrr @{}}
\toprule
 & & \multicolumn{3}{c}{Bedroom} & \multicolumn{3}{c}{Dining room} & \multicolumn{3}{c}{Living room} \\
\cmidrule(l{0pt}r{2pt}){3-5} \cmidrule(l{0pt}r{2pt}){6-8} \cmidrule(l{0pt}r{2pt}){9-11}
Coloring & Method & \fid & \kid & \sca & \fid & \kid & \sca & \fid & \kid & \sca  \\
\midrule
\multirow{3}{*}{\shortstack[c]{Semantic\\DiffuScene palette}} &
\atiss & 71.58 & 5.65 & 87.56 & 59.33 & 13.22 & 82.07 & 57.84 & 15.35 & 80.39 \\
& \diffuscene & 32.98 & 0.49 & 54.16 & 45.79 & 0.86 & 49.37 & 45.37 & 0.81 & 51.67 \\
& \layoutgpt & 71.33 & 4.81 & 98.73 & 57.51 & 13.47 & 71.93 & 59.74 & 14.78 & 79.90 \\
\midrule
\multirow{3}{*}{\shortstack[c]{Semantic\\Our new palette}} &
\atiss & 70.68 & 24.87 & 94.45 & 72.79 & 21.35 & 90.73 & 78.33 & 24.88 & 91.44 \\
& \diffuscene & 34.52 & 1.12 & 58.01 & 49.69 & 0.51 & 49.86 & 50.10 & 0.45 & 53.41 \\
& \layoutgpt & 57.73 & 23.33 & 93.08 & 57.64 & 8.35 & 84.24 & 60.96 & 8.38 & 83.03 \\
\midrule
\multirow{3}{*}{\shortstack[c]{Textured}} &
\atiss & 83.12 & 2.21 & 57.02 & 77.69 & 8.72 & 74.02 & 72.61 & 10.72 & 67.29 \\
& \diffuscene & 46.55 & 0.34 & 50.43 & 63.03 & 0.39 & 44.43 & 60.27 & 0.47 & 48.75 \\
& \layoutgpt & 100.85 & 17.34 & 99.70 & 115.34 & 36.90 & 95.61 & 91.86 & 28.13 & 99.89 \\
\bottomrule
\end{tabular}
}
\label{tab:check-colour-palette}
\end{table*}

\begin{table*}
\centering
\caption{
Comparison of rendering with simple-3dviz vs our rendering tool (STK) in different settings, which supports rendering of arbitrary floor shapes. Scenes are generated without any conditioning using official pretrained models (separate models are used for each room type).  As original \diffuscene~\cite{tang2023diffuscene} cannot generate with floor conditioning, we use a square floor for the generated scenes.  We compare rendering the GT with the actual floorplan or the square floor, and rendering with different scales.
}
\vspace{-8pt}
\resizebox{\linewidth}{!}
{
\begin{tabular}{@{} ccc rrr rrr rrr @{}}
\toprule
 & & & \multicolumn{3}{c}{Bedroom} & \multicolumn{3}{c}{Dining room} & \multicolumn{3}{c}{Living room}
 \\
\cmidrule(l{0pt}r{2pt}){4-6} \cmidrule(l{0pt}r{2pt}){7-9} \cmidrule(l{0pt}r{2pt}){10-12}
GT & Renderer & Scale & \fid & \kid & \sca  & \fid & \kid & \sca  & \fid & \kid & \sca \\
\midrule
square & simple-3dviz & Fixed & 
34.52 & 1.12 & 58.01 & 49.69 & 0.51 & 49.86 & 50.10 & 0.45 & 53.41  \\
\midrule
\multirow{2}{*}{square} & \multirow{2}{*}{STK} & Fixed & 68.01 & 1.45 & 71.50 & 45.18 & 3.33 & 66.09 & 44.27 & 5.83 & 71.11
 \\
& & Zoomed-in & 58.80 & 5.08 & 86.67 & 41.96 & 4.93 & 85.67 & 39.34 & 5.88 & 82.42 \\
\midrule
\multirow{2}{*}{actual} & \multirow{2}{*}{STK} & Fixed & 75.79 & 15.65 & 100.00 & 79.86 & 41.82 & 99.99 & 77.41 & 44.78 & 99.99
 \\
& & Zoomed-in & 100.90 & 47.96 & 99.94 & 141.03 & 108.74 & 98.35 & 121.78 & 103.92 & 99.19 \\
\bottomrule
\end{tabular}
}
\label{tab:check-render-tool}
\end{table*}

\label{sec:supp-prelim-experiments}

For our experiments, we use a modified color palette and rendering setup differing from prior work~\cite{paschalidou2021atiss,tang2023diffuscene,hu2024mixed}.  Our modified palette (\cref{fig:vis-color-palette}) was selected to be a single palette that included all object categories across the different room types, and used similar colors for closely related object categories (e.g. greens for the different beds -- \semtype{kids bed, single bed, double bed}).  For rendering, we developed our own tool to render an arbitrarily shaped room layout with architecture information (e.g. showing the placement of windows and doors).  

We render with a flatter shading, whereas earlier work used a camera‑mounted point light in \texttt{\small simple3dviz}. Compared with prior work, we also render with a more zoomed-in view that allows for closer inspection of the object details. 
\Cref{fig:vis-diff-color-qua} shows examples comparing renderings using our rendering tool (STK) and our new color palette and those using simple3dviz with the color palette used by DiffuScene and prior work.
\Cref{fig:vis-color-palette} shows the differences in color palettes between prior work (left two) and ours (right). Our color palette assigns similar colors to object categories with similar functions.
We conduct a series of experiments to account for how changes in rendering styles impact the common evaluation metrics used in scene generation.
We first conduct initial sanity experiments on DiffuScene~\cite{tang2023diffuscene} (\cref{sec:supp-expr-reproduce-prior-work}), to set a point of comparison for our studies on the impact of changes in renderings (\cref{sec:supp-expr-rendering}).

\subsection{Reproducing prior work results}
\label{sec:supp-expr-reproduce-prior-work}

We used the pretrained \diffuscene weights for the sanity check.
The training is on a single A40 GPU with a batch size of 128 for $100k$ epochs.
We set the learning rate $l_r=2e^{-4}$ with a decay rate of $0.5$ in every $15k$ epochs.
The noise intensity linearly increases from $0.0001$ to $0.02$ with $1000$ time steps.
The above settings exactly follow the original implementation of \diffuscene.
However, from \Cref{tab:diffuscene-baseline}, the pretrained results are not as good as reported in the paper \cite{tang2023diffuscene}.
We also tried to retrain the DiffuScene model to follow the DiffuScene instruction and setting exactly.
We find that both the retrained and pretrained models perform worse than the reported results in the paper.%

To confirm that our DiffuScene model is trained well, we use a rendering similar to the one used in the original paper (semantic class-based rendering without floor) and compare the results with those reported in the paper (\Cref{tab:diffuscene-baseline}).  We find that our retrained model has better KID and SCA than reported in the paper, but slightly worse FID and CKL.  Via correspondence with the authors, we confirmed that some deviation from the reported numbers in the paper is expected when retraining.

\vspace{-5pt}
\subsection{Impact of rendering style on metrics}
\label{sec:supp-expr-rendering}
\vspace{-5pt}
We note that the metrics are strongly influenced by rendering style, including whether a floor is included in the rendering, whether textures or semantic coloring are used, and the precise semantic coloring palette used.  To check the impact of each of these choices on the evaluation metrics, we conduct a series of experiments.  For these experiments, we generate and save a thousand scenes (for each room-type) for each method with official pretrained checkpoints, and then vary the rendering style based on the experiment.  Thus, the scenes are kept constant with just the rendering style being changed.  

Experiments here use \emph{per-roomtype} models (i.e. separate models trained for bedroom, dining room, and living room) except for LayoutGPT.

\mypara{Impact of rendering style.} Prior work has used different types of renderings for evaluation, including top-down semantic classes~\cite{tang2023diffuscene}, top-down textured~\cite{paschalidou2021atiss}, and perspective textured~\cite{feng2023layoutgpt}.  
In our experiments, we use the top-down birds-eye-view as it provides a more holistic view of the scene.  
Prior work also typically does not include the floor and architecture elements as part of the rendering.  As our work focuses on whether the objects are correctly placed on the floor, we compare how the evaluation metrics change depending on whether the floor is included in the rendering or not.  For this initial investigation of the rendering style, we use a square floor (filling up the entire image) to see how much including the floor impacts the evaluation metrics.  

We also notice that the semantic coloring used by DiffuScene does not visually group similar categories together.  Thus, we also compare the impact of using a different semantic color palette for evaluation. 
In our revised color palette, we choose similar colors for object categories that are semantically closer together (e.g. can be grouped into a coarser category).  See \cref{fig:vis-diff-color-qua} for examples of scenes rendered with the new color palette vs the original color palette, and \cref{fig:vis-color-palette} for the correspondence between colors and semantic object classes.

As some prior work used textured assets, we also investigate how using textured vs semantic top-down rendering affects the metrics for DiffuScene in \Cref{tab:check-textured-vs-semantic-diffuscene}.  We report only FID, KID, and SCA, as these are the metrics impacted by the rendering style. 
We see from \cref{tab:check-textured-vs-semantic-diffuscene} that the view-based metrics are indeed affected, with higher FID and lower KID when a square floor is included.

\begin{figure*}[ht]
\includegraphics[width=\textwidth]{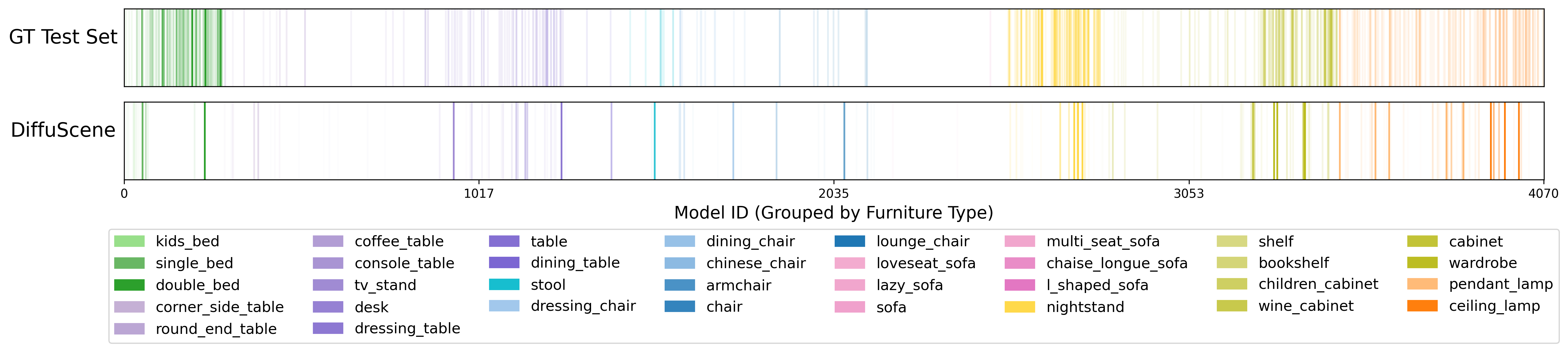}
\caption{
Model-ID frequency histograms for bedroom.
Color intensity indicates how often each model ID (specific object model) appears in the ground-truth test set (top) and in scenes generated by DiffuScene (bottom). Model IDs are grouped by object category, and each category is shown in a distinct colour. 
The ground truth 3DFront scenes distribute their counts over many model IDs, whereas DiffuScene uses only a small subset, revealing lower shape diversity in its generated results.
}
\vspace{-4pt}
\label{fig:vis-model-id}
\end{figure*}

\mypara{Does rendering style affect evaluation of methods in prior work?} In ~\Cref{tab:check-colour-palette}, we check how the semantic colorings and using textures affect the ordering of different methods.  We compare ATISS~\cite{paschalidou2021atiss}, DiffuScene~\cite{tang2023diffuscene}, and LayoutGPT~\cite{feng2023layoutgpt}.

\emph{Colour palette for semantic rendering.} With our revised semantic palette, we find LayoutGPT performance to be more clearly separated (and comparatively better) than that of ATISS.

\emph{Textured vs semantic rendering.} Overall, with semantic renderings, the FID is consistently lower and the KID consistently higher.  With the FID and KID, the relative rankings of the methods are the same.

\begin{table*}
\centering
\caption{
Accuracy of orientation baselines with 4 classes on z-axis direction with different error thresholds for different baselines. 
}
\vspace{-8pt}
\small
{
\begin{tabular}{@{} c rrr rrr rrr | rrr @{}}
\toprule
 & \multicolumn{3}{c}{Bedroom} & \multicolumn{3}{c}{Living room} & \multicolumn{3}{c}{Dining room} 
 & \multicolumn{3}{c}{Overall (average)} 
 \\
\cmidrule(l{0pt}r{2pt}){2-4} \cmidrule(l{0pt}r{2pt}){5-7} \cmidrule(l{0pt}r{2pt}){8-10} \cmidrule(l{0pt}r{2pt}){11-13}
& 
$<5^{\circ}$ & $<10^{\circ}$ & $<45^{\circ}$ & 
$<5^{\circ}$ & $<10^{\circ}$ & $<45^{\circ}$ & 
$<5^{\circ}$ & $<10^{\circ}$ & $<45^{\circ}$ & 
$<5^{\circ}$ & $<10^{\circ}$ & $<45^{\circ}$  \\
\midrule
Random & 23.7 & 24.3 & 25.0 & 23.6 & 23.6 & 24.6  & 23.1 & 23.1 & 23.5 & 23.5	& 23.7 & 24.4 \\
Majority & 38.2 & 38.3 & 39.3 & 34.4 & 34.4 & 36.1 & 34.9 & 34.9 & 36.8 & 35.8	& 35.9 & 37.4 \\
Inward & 56.4 & 56.5 & 56.8 & 34.4 & 34.5 & 36.3 & 31.8 & 31.8 & 33.7 & 40.9	& 40.9 & 42.3 \\
APM & 81.5 & 81.5 & 82.7 & 77.1 & 77.2 & 80.2 & 76.5 & 76.6 & 79.9 & 78.4	& 78.4 & 80.9\\
\midrule
GT & 96.2 & 96.4 & 100.0 & 94.1 & 94.3 & 98.8 & 94.1 & 94.3 & 98.8 & 94.8	& 95.0 & 99.2 \\
\bottomrule
\end{tabular}
}
\label{tab:orientation-baselines}
\end{table*}

\mypara{Comparing rendering with simple-3dviz vs STK.}
We also compare rendering with a square floorplan vs the actual floorplan.  To allow for rendering with different floor shapes, we develop our own rendering tool that can generate arbitrary floor geometry given the set of corner points.  
\Cref{tab:check-render-tool} reports semantic results with our new color palette only.
Switching from simple-3dviz to our STK renderer under the same fixed scale leaves FID/KID almost unchanged but raises SCA slightly.
Changing the camera to the zoomed-in view (no padding) increases SCA further because larger projected objects are easier to classify, while FID/KID shift only marginally.
When the reference images are rendered with the actual, non-square floor shape, FID and KID climb steeply for both scales—an expected penalty because the generated scenes still assume a square outline and thus diverge in floor-pixel distribution.

\emph{Impact on SCA.} Intriguingly, we find the SCA dramatically increases when we switched to our updated rendering settings, indicating that under this setting, it was easy for the classifier to distinguish between generated and ground-truth scenes.

\subsection{SCA analysis}
\label{sec:supp-sca-analysis}
Upon investigating the change in SCA further, we find that the retrieval process results in very limited set of objects being selected for generated scenes, causing the classifier to easily identify which scenes are generated.  This was not apparent with the rendering choices in prior work, as the objects are rendered too small and without sufficient details.

We test this hypothesis by plotting a histogram of 3D model IDs used in ground truth vs generated scenes.  From \Cref{fig:vis-model-id}, we see that ground-truth bedrooms use a diverse set of object models, whereas DiffuScene relies on a narrow subset, amplifying appearance differences that STK makes more visible. These differences come from the object-retrieval step, which cannot guarantee a similar object model distribution, and show that SCA is an unreliable metric when renderings vary.

\textit{SCA using model ID.} We also run a simple test experiment to investigate the impact of the model ID on the SCA. Instead of rendered image, we use the set of model ID in the scene to represent the scene. More specifically, we represent scenes using a bag-of-words approach, creating binary vectors where each dimension indicates the presence of a specific furniture model ID. For classification, we use a simple neural network with one hidden layer that learns to distinguish between real and synthetic scenes based on the exact model IDs present in each scene.
Our model achieved approximately $99\%$ accuracy when trained on a dataset comprising ground truth training scenes and DiffuScene results, then evaluated on ground truth test scenes and separate DiffuScene results. To verify the model's reliability, we conducted a control experiment by randomizing the splits of only ground truth data for both training and testing. This control test resulted in accuracy around $50\%$, equivalent to random guessing, confirming our classifier properly learns meaningful distinctions rather than exploiting dataset artifacts.

\begin{table*}[t]
\centering
\caption{
Evaluation of how well scenes generated using different methods match the distribution of ground-truth scenes for different room types. 
We report the \fid, \kid, \sca, and \ckl with different levels of architecture conditioning: unconditioned (None), conditioned on the floor (Floor), and the architecture map (Arch). 
\textbf{Bold} indicates best results.
Our \semdiff performs better than other methods in most evaluation metrics across different room types and conditions.}
\vspace{-8pt}
\resizebox{\linewidth}{!}
{
\begin{tabular}{@{} l c rrrr rrrr rrrr | rrrr @{}}
\toprule
 & & \multicolumn{4}{c}{Bedroom} & \multicolumn{4}{c}{Dining room} & \multicolumn{4}{c}{Living room} & \multicolumn{4}{c}{Overall (average)} \\
\cmidrule(l{0pt}r{2pt}){3-6} \cmidrule(l{0pt}r{2pt}){7-10} \cmidrule(l{0pt}r{2pt}){11-14} \cmidrule(l{0pt}r{2pt}){15-18}
Condition & Method & \fid & \kid & \sca & \ckl & \fid & \kid & \sca & \ckl & \fid & \kid & \sca & \ckl & \fid & \kid & \sca & \ckl \\
\midrule
\multirow{4}{*}{\shortstack[l]{None}} 
& \diffuscene \cite{tang2023diffuscene} & 
{113.85} & {61.28} & {99.26} & 60.89 & 
{137.38} & {97.16} & 100.00 & \textbf{3.66} & 
{125.15} & {105.82} & 100.00 & {9.55} & 
{125.46} & {88.09} & {99.75} & 24.70 \\
& \midiff \cite{hu2024mixed} & 
114.97 & 52.49 & 96.88 & 62.38 & 
107.28 & 46.19 & 98.50 & 21.17 & 
79.38 & 25.13 & 96.66 & \textbf{0.69} & 
100.54 & 41.27 & 97.35 & 28.31 \\
& \semdiff (Ours) & 
\textbf{103.27} & \textbf{9.80} & \textbf{98.59} & \textbf{31.99} & 
\textbf{92.70} & \textbf{11.33} & \textbf{92.27} & {9.52} & 
\textbf{85.82} & \textbf{11.02} & \textbf{99.41} & 10.13 & 
\textbf{93.93} & \textbf{10.72} & \textbf{96.76} & \textbf{17.21}\\
\midrule
\multirow{4}{*}{\shortstack[l]{Floor}} 
& \diffuscene \cite{tang2023diffuscene} & 
{103.14} & {44.75} & {94.83} & 61.42 & 
{94.79} & {26.09} & {91.40} & \textbf{2.33} & 
{74.52} & {19.55} & {97.08} & 8.04 & 
{90.82} & {30.13} & {94.43} & 23.93  \\
& \midiff \cite{hu2024mixed} & 106.81 & 34.39 & 96.88 & {48.60} & 
94.11 & 24.18 & 98.51 & 21.18 & 
74.45 & 20.10 & 96.42 & \textbf{1.13} & 
91.79 & 26.23 & 97.27 & {23.64} \\
& \semdiff (Ours) & 
\textbf{84.93} & \textbf{12.61} & \textbf{93.05} & \textbf{10.08} & 
\textbf{87.44} & \textbf{16.72} & \textbf{84.92} & {4.25} & 
\textbf{73.01} & \textbf{14.23} & \textbf{91.51} & {3.63} & 
\textbf{81.79} & \textbf{14.52} & \textbf{89.83} & \textbf{5.99} \\
\midrule
\multirow{3}{*}{\shortstack[l]{Arch}} 
& \diffuscene \cite{tang2023diffuscene} & 
{103.16} & 47.46 & {95.49} & {40.89} & 
87.45 & 21.14 & 94.51 & 8.92 & 
74.79 & 22.06 & 95.07 & 11.47 & 
88.47 & 30.22 & 95.02 & 20.42 \\
& \midiff \cite{hu2024mixed} & 
103.74 & 32.06 & 94.58 & 61.72 & 
95.93 & 36.47 & 97.17 & 43.71 & 
80.87 & 24.66 & 94.08 & 5.48 & 
93.51 & 31.06 & 95.28 & 36.97 \\
& \semdiff (Ours) & 
\textbf{75.42} & \textbf{5.93} & \textbf{86.81} & \textbf{5.27} & 
\textbf{70.87} & \textbf{9.97} & \textbf{85.79} & \textbf{4.97} & 
\textbf{66.90} & \textbf{10.05} & \textbf{88.29} & \textbf{4.10} & 
\textbf{71.06} & \textbf{8.65} & \textbf{86.96} & \textbf{4.78} \\ 
\bottomrule
\end{tabular}
}
\label{tab:quant-distmatch-full}
\end{table*}
\begin{table*}
\centering
\caption{
Evaluation of the physical plausibility of the generated scenes for different room types. We compare out-of-bounds (OOB) ratios at the scene (\oobscene) and object level (\oobobj), as well as the collision rate (\col), and navigability (\nav) for the architecture-conditioned model. 
}
\vspace{-8pt}
\resizebox{\linewidth}{!}
{
\begin{tabular}{@{} c rrrr rrrr rrrr | rrrr @{}}
\toprule
 & \multicolumn{4}{c}{Bedroom} & \multicolumn{4}{c}{Dining room} & \multicolumn{4}{c}{Living room} & \multicolumn{4}{c}{Overall (average)} \\
 \cmidrule(l{0pt}r{2pt}){2-5} 
 \cmidrule(l{0pt}r{2pt}){6-9} 
 \cmidrule(l{0pt}r{2pt}){10-13}
 \cmidrule(l{0pt}r{2pt}){14-17}
Method & 
\oobscene & \oobobj & \col & \nav &
\oobscene & \oobobj & \col & \nav &
\oobscene & \oobobj & \col & \nav &
\oobscene & \oobobj & \col & \nav
\\
\midrule
\diffuscene \cite{tang2023diffuscene} & 
55.10 & 20.77 & 37.69 & 91.98 &
71.50 & 30.78 & 42.65 & 96.07 &
72.80 & 26.53 & 42.57 & 94.98 &
66.47 & 20.03 & 40.97 & 94.34
\\
\midiff & 
65.25 & 37.81 & 33.60 & \textbf{95.25} &
46.63 & 30.74 & 64.62 & 97.22 & 
70.15 & 35.44 & 56.64 & \textbf{96.10} &
60.68	& 34.66	&  51.62 & 96.19
\\
\semdiff (Ours) & 
\textbf{13.80} & \textbf{4.00} & \textbf{26.21} & 93.80 &
\textbf{15.80} & \textbf{7.53} & \textbf{5.37} & \textbf{99.36} &
\textbf{20.30} & \textbf{8.32} & \textbf{16.82} & 95.92 &
\textbf{16.63} & \textbf{6.62} & \textbf{16.13} & \textbf{96.36}
\\
\bottomrule
\end{tabular}
\vspace{-1em}
}
\label{tab:quant-plausibility-full}
\end{table*}

\begin{table}
\centering
\caption{
Statistics on the number of objects across different room types with architectural conditioning results.
}
\vspace{-8pt}
\resizebox{\linewidth}{!}
{
\begin{tabular}{@{} c rrrr @{}}
\toprule
 & Bedroom & Dining room & Living room & Overall \\
\midrule
\diffuscene & 5.03 & 10.97  & 9.95 & 8.65\\
\midiff & 5.11 & 7.31 & 9.63 & 7.35\\ 
\semdiff & 4.65 & 8.23 & 9.47 & 7.45\\
\midrule
GT (train) & 4.96 & 10.71 & 11.51 & 6.40 \\
GT (test) & 5.09 & 11.06 & 11.49 & 9.39 \\
\bottomrule
\end{tabular}
}
\label{tab:objstats}
\end{table}

\begin{table*}
\centering
\caption{
Evaluation directly on layouts. Results are evaluated on architecture-conditioned generated scenes using semantic 3D bounding boxes, without performing object retrieval. \semdiff significantly outperforms other prior methods in the pure layout.
}
\vspace{-8pt}
\resizebox{\linewidth}{!}
{
\begin{tabular}{@{} l c rrr rrr rrr | rrr @{}}
\toprule
 & & \multicolumn{3}{c}{Bedroom} & \multicolumn{3}{c}{Dining room} & \multicolumn{3}{c}{Living room} & \multicolumn{3}{c}{Overall (average)} \\
\cmidrule(l{0pt}r{2pt}){3-5} \cmidrule(l{0pt}r{2pt}){6-8} \cmidrule(l{0pt}r{2pt}){9-11} \cmidrule(l{0pt}r{2pt}){12-14}
Condition & Method & \fid & \kid & \sca & \fid & \kid & \sca & \fid & \kid & \sca & \fid & \kid & \sca \\
\midrule
\multirow{3}{*}{Arch} &
\diffuscene \cite{tang2023diffuscene} & 
87.91 & 100.47 & 99.99 & 93.74 & 
94.59 & 98.91 & 98.63 & 114.35 & 
100.00 & 93.43 & 103.14 & 99.63 \\
& \midiff \cite{hu2024mixed} & 
73.54 & 78.79 & 95.72 & 71.95 & 
74.83 & 97.31 & 80.89 & 83.74 & 
98.16 & 75.46 & 79.12 & 97.06 \\
& \semdiff (Ours) & 
\best{24.16} & \best{14.85} & \best{94.58} & \best{27.46} & 
\best{10.75} & \best{93.49} & \best{26.88} & \best{16.74} & 
\best{96.48} & \best{26.17} & \best{14.11} & \best{94.85} \\
\bottomrule
\end{tabular}
}
\label{tab:quant-bbox}
\end{table*}

\begin{table*}[t]
\centering
\caption{
Evaluation (with breakdown for each room type) of distribution match for \semdiff under single condition (per-masktype) training and mixed condition training.
}
\vspace{-8pt}
\resizebox{\linewidth}{!}
{
\begin{tabular}{@{} l c rrrr rrrr rrrr | rrrr @{}}
\toprule
 & & \multicolumn{4}{c}{Bedroom} & \multicolumn{4}{c}{Dining room} & \multicolumn{4}{c}{Living room} & \multicolumn{4}{c}{Overall (average)} \\
\cmidrule(l{0pt}r{2pt}){3-6} \cmidrule(l{0pt}r{2pt}){7-10} \cmidrule(l{0pt}r{2pt}){11-14} \cmidrule(l{0pt}r{2pt}){15-18}
Condition & Training & \fid & \kid & \sca & \ckl & \fid & \kid & \sca & \ckl & \fid & \kid & \sca & \ckl & \fid & \kid & \sca & \ckl \\
\midrule
\multirow{2}{*}{\shortstack[l]{None}} 
& per-masktype & 
103.27 & \best{9.80} & 98.59 & 31.99 & 
92.70 & \best{11.33} & \best{92.27} & 9.52 & 
85.82 & \best{11.02} & 99.41 & \best{10.13} & 
93.93 & \best{10.72} & 96.76 & 17.21\\
& mixed & 
\best{76.81} & 10.39 & \best{91.20} & \best{6.15} & 
\best{77.05} & \best{12.14} & 93.69 & \best{6.11} & 
\best{63.67} & 11.27 & \best{93.54} & \best{3.85} & 
\best{72.51} & 11.27 & \best{92.81} & \best{5.37} \\
\midrule
\multirow{2}{*}{\shortstack[l]{Floor}} 
& per-masktype & 
84.93 & 12.61 & 93.05 & 10.08 & 
87.44 & 16.72 & 84.92 & 4.25 & 
73.01 & 14.23 & 91.51 & 3.63 & 
81.79 & 14.52 & 89.83 & 5.99 \\
& mixed & 
\best{79.24} & \best{10.01} & \best{88.00} & \best{6.20} & 
\best{80.11} & \best{10.89} & \best{83.76} & \best{4.03} & 
\best{69.37} & \best{10.89} & \best{86.35} & \best{2.92} & 
\best{76.24} & \best{10.60} & \best{86.04} & \best{4.38} \\
\midrule
\multirow{2}{*}{\shortstack[l]{Arch}} 
& per-masktype & 
\best{75.42} & \best{5.93} & \best{86.81} & \best{5.27} & 
\best{70.87} & 9.97 & \best{85.79} & 4.97 & 
66.90 & 10.05 & \best{88.29} & 4.10 & 
\best{71.06} & \best{8.65} & \best{86.96} & 4.78 \\ 
& mixed & 
78.95 & 11.10 & 92.29 & 6.37 & 
76.53 & \best{8.72} & 89.95 & \best{3.69} & 
\best{65.33} & \best{9.72} & 94.32 & \best{2.77} & 
73.61 & 9.85 & 92.19 & \best{4.27}\\
\bottomrule
\end{tabular}
}
\label{tab:quant-distmatch-mix}
\end{table*}
\begin{table*}
\centering
\caption{
Evaluation (with breakdown for each room type) of the physical plausibility of the generated scenes for \semdiff under single condition (per-masktype) training and mixed condition training.
}
\vspace{-8pt}
\resizebox{\linewidth}{!}
{
\begin{tabular}{@{} l c rrrr rrrr rrrr | rrrr @{}}
\toprule
& & \multicolumn{4}{c}{Bedroom} & \multicolumn{4}{c}{Dining room} & \multicolumn{4}{c}{Living room} & \multicolumn{4}{c}{Overall (average)} \\
 \cmidrule(l{0pt}r{2pt}){3-6} \cmidrule(l{0pt}r{2pt}){7-10} \cmidrule(l{0pt}r{2pt}){11-14}
 \cmidrule(l{0pt}r{2pt}){15-18}
Condition & Method & 
\oobscene & \oobobj & \col & \nav &
\oobscene & \oobobj & \col & \nav &
\oobscene & \oobobj & \col & \nav &
\oobscene & \oobobj & \col & \nav 
\\
\midrule
\multirow{2}{*}{\shortstack[l]{None}} &
per-masktype & 
4.04 & 1.13 & 24.72 & 93.02 & 
1.04 & \textbf{0.13} & 14.94 & 96.36 & 
1.04 & 0.13 & 13.45 & 96.52 & 
2.04 & 0.46 & 17.70 & 95.30 \\
& mixed & 
\textbf{1.30} & \textbf{0.36} & \textbf{19.02} & \textbf{95.70} & 
\textbf{0.61} & 0.15 & \textbf{11.90} & \textbf{97.07} & 
\textbf{0.20} & \textbf{0.02} & \textbf{12.04} & \textbf{97.74} & 
\textbf{0.70} & \textbf{0.18} & \textbf{14.32} & \textbf{96.84}  \\

\midrule
\multirow{2}{*}{\shortstack[l]{Floor}} &  
per-masktype & 
12.80 & 4.55 & \textbf{17.84} & \textbf{95.51} & 
28.90 & 7.51 & \textbf{11.65} & \textbf{97.98} & 
32.10 & 8.73 & 29.34 & 96.57 & 
24.60 & 6.93 & 19.61 & \textbf{96.69} \\
& mixed & 
\textbf{9.60} & \textbf{3.32} & 19.60 & 95.08 & 
\textbf{25.30} & \textbf{5.57} & 11.99 & 97.04 & 
\textbf{29.60} & \textbf{6.31} & \textbf{12.41} & \textbf{97.78} & 
\textbf{21.50} & \textbf{5.07} & \textbf{14.67} & 96.63 \\

\midrule
\multirow{2}{*}{\shortstack[l]{Arch}} &  
per-masktype & 
13.80 & 4.00 & 26.21 & 93.80 &
\textbf{15.80} & 7.53 & \textbf{5.37} & \textbf{99.36} & 
\textbf{20.30} & 8.32 & 16.82 & 95.92 & 
\textbf{16.63} & 6.62 & 16.13 & 96.36  \\
& mixed & 
\textbf{10.40} & \textbf{3.46} & \textbf{18.48} & \textbf{95.86} &
26.10 & \textbf{5.75} & 12.60 & 97.37 &
28.10 & \textbf{5.73} & \textbf{12.19} & \textbf{97.49} &
21.53 & \textbf{4.98} & \textbf{14.42} & \textbf{96.91}  \\
\bottomrule
\end{tabular}
\vspace{-1em}
}
\label{tab:quant-plausibility-mix}
\end{table*}

\begin{table*}
\centering
\caption{
Evaluation of how well scenes generated using different methods match the distribution of ground-truth scenes for \textit{per-roomtype} models conditioned on architecture plan. 
}
\vspace{-8pt}
\resizebox{\linewidth}{!}
{
\begin{tabular}{@{} c rrrr rrrr rrrr | rrrr@{}}
\toprule
& \multicolumn{4}{c}{Bedroom} & \multicolumn{4}{c}{Dining room} & \multicolumn{4}{c}{Living room} & \multicolumn{4}{c}{Overall (average)}\\
\cmidrule(l{0pt}r{2pt}){2-5} \cmidrule(l{0pt}r{2pt}){6-9} \cmidrule(l{0pt}r{2pt}){10-13} \cmidrule(l{0pt}r{2pt}){14-17}
Method & \fid & \kid & \sca & \ckl & \fid & \kid & \sca & \ckl & \fid & \kid & \sca & \ckl & \fid & \kid & \sca & \ckl \\
\midrule
\diffuscene \cite{tang2023diffuscene} & 148.56 & 75.68 & 99.61 & 22.48 & 88.83 & 20.89 & 95.42 & \best{11.38} & \best{73.70} & \best{17.37} & 94.75 & \best{11.00} & \best{103.70} & \best{37.98} & 96.59 & 82.41 \\
\midiff \cite{hu2024mixed} & \best{106.32} & \best{69.72} & 99.80 & 14.46 & 135.20 & 92.80 & 99.03 & 58.90 & 89.87 & 36.43 & 99.71 & 59.15 & 110.46 & 66.31 & 99.52 & 44.17 \\
\semdiff (Ours) & 156.69 & 85.00 & 99.07 & \best{7.05} & \best{86.72} & \best{20.14} & 95.48 & 12.13 & 80.57 & 23.84 & 93.70 & 30.53 & 107.99 & 42.99 & 96.08 & \best{16.57} \\ 
\bottomrule
\end{tabular}
}
\label{tab:quant-distmatch-single-arch}
\end{table*}
\begin{table*}
\centering
\caption{
Evaluation of the physical plausibility of the generated scenes for per-roomtype models. We compare out-of-bounds (OOB) ratios  at the scene (\oobscene) and object level (\oobobj), as well as the collision rate (\col), and navigability (\nav) for the architecture-conditioned model.}
\vspace{-8pt}
\resizebox{\linewidth}{!}
{
\begin{tabular}{@{} c rrrr rrrr rrrr | rrrr @{}}
\toprule
 & \multicolumn{4}{c}{Bedroom} & \multicolumn{4}{c}{Dining room} & \multicolumn{4}{c}{Living room} & \multicolumn{4}{c}{Overall (average)} \\
 \cmidrule(l{0pt}r{2pt}){2-5} \cmidrule(l{0pt}r{2pt}){6-9} \cmidrule(l{0pt}r{2pt}){10-13}
 \cmidrule(l{0pt}r{2pt}){14-17}
Method & 
\oobscene & \oobobj & \col & \nav &
\oobscene & \oobobj & \col & \nav &
\oobscene & \oobobj & \col & \nav &
\oobscene & \oobobj & \col & \nav 
\\
\midrule
\diffuscene \cite{tang2023diffuscene} & 
57.00 & 21.66 & \textbf{21.61} & 91.18 &
71.90 & 35.05 & 37.96 & 95.70 & 
76.10 & 33.75 & 29.66 & 95.51 &
68.33 & 30.15 & 29.74 & 94.13 \\
\midiff \cite{hu2024mixed} & 
15.50  & 11.72 & 68.56 & \textbf{97.69} &
\textbf{14.50} & 28.82 & 72.13 & \textbf{99.23} & 
\textbf{32.00} & 17.92 & 82.68 & \textbf{97.39} &
\textbf{20.67}	& 19.49	&  74.46 & \textbf{98.10} \\
\semdiff (Ours) & 
\textbf{14.30} & \textbf{4.47} & 22.61 & 97.11 &
34.00 & \textbf{12.15} & \textbf{20.22} & 94.12 &
32.60 & \textbf{10.04} & \textbf{26.53} & 94.10 &
26.97 & \textbf{8.89} & \textbf{23.11} & 95.11 \\
\bottomrule
\end{tabular}
\vspace{-1em}
}
\label{tab:quant-plausibility-single-arch}
\end{table*}

\begin{table*}
\centering
\caption{
Comparison with PhyScene for the living room with floor condition. Most of the scenes generated using PhyScene have out-of-bounds (OOB) issues, indicating that the objects are not placed within the provided floor mask.
}
\vspace{-8pt}
{
\begin{tabular}{@{} c rrrr|rrrr@{}}
\toprule
& \multicolumn{4}{c}{Distribution} & \multicolumn{4}{c}{Plausibility} \\
\cmidrule(l{0pt}r{2pt}){2-5} \cmidrule(l{0pt}r{2pt}){6-9} Method & \fid & \kid & \sca & \ckl & \oobscene & \oobobj & \col & \nav \\
\midrule
PhyScene~\cite{yang2024physcene} & 73.50 & 13.30 & 95.52 & \best{0.99} & 84.50 & 38.86 & 34.44 & 95.54 \\
\semdiff & \best{68.26} & \best{7.65} & \best{92.13} & 3.52 & \best{26.50} & \best{6.53} & \best{24.57} & \best{96.24} \\
\bottomrule
\end{tabular}
}
\label{tab:quant-physcene}
\end{table*}

\section{Additional results}
\label{sec:supp-additional-results}

In this section, we present additional results, including an evaluation of our attribute predictor  (\cref{sec:supp-expr-attr-pred}),
 additional results for our main model (\cref{sec:supp-expr-per-archtype}),
 direct evaluation of the predicted layout (\cref{sec:supp-expr-layout}), 
 details of mixed condition training (\cref{sec:supp-mix}),
 results for per-room type models (\cref{sec:supp-expr-per-room-model}), 
 comparison against PhyScene (\cref{sec:supp-physcene}), discussion of failure cases  (\cref{sec:supp-failure-cases}),
 and additional examples of generated scenes from our \semdiff (\cref{sec:supp-expr-qualitative-results}).

\subsection{Attribute prediction}
\label{sec:supp-expr-attr-pred}

\mypara{Baselines for orientation prediction.}
We compare our attribute prediction model (APM) orientation prediction with several baselines.  We note that most objects are oriented parallel to the room coordinate axes, so we simplify the orientation prediction as a classification problem.  We compare using a random baseline, vs the most frequent orientation (across all object categories and conditioned on the object category), as well as an inward-facing heuristic.  We measure the accuracy of the predicted orientation under different error thresholds.  Our trained APM achieves the best performance and outperforms baselines (\cref{tab:orientation-baselines}).

\mypara{Vertical position and object height prediction.}
We also evaluate our APM for the vertical position (offset from floor) and object height on the unified test set by computing the mean absolute error (MAE) between the predicted and GT offset and height.  
We get the offset MAE of 0.13m, and the height MAE of 0.15m.
The results show that APM can perform well on the 3D attributes given the 2D info.

\begin{figure}[ht]
\includegraphics[width=\linewidth]{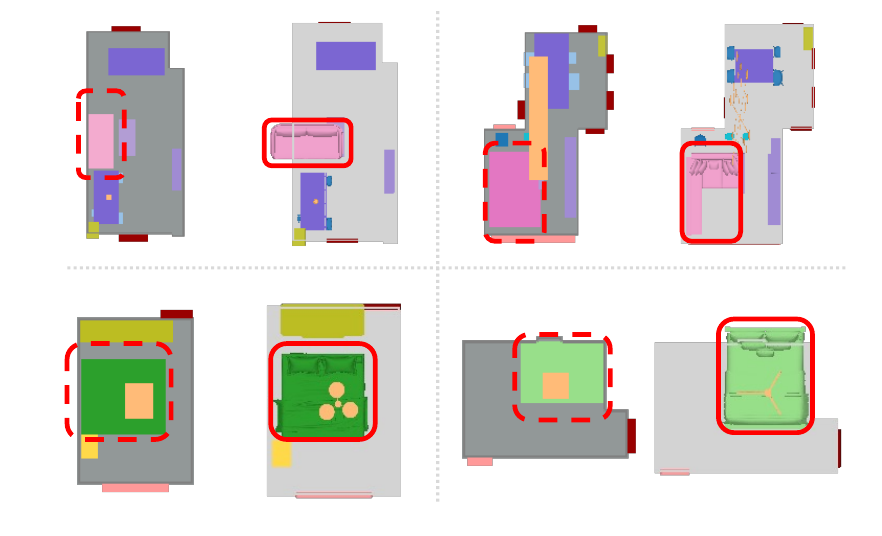}
\caption{Failure cases due to poor object attribute prediction and retrieval with our \semdiff arch plan conditioning. 
For each scene, we show the generated semantic map (left) and the final scene after object retrieval (right).
On the semantic map, the box (with dashed lines) highlights an object for which the attribute prediction was inaccurate. For the final scene, the red solid line shows the retrieved object (typically with incorrect orientation) due to poor attribute prediction.
}
\vspace{-4pt}
\label{fig:vis-qua-bad-attr}
\end{figure}

\mypara{Limitations in object attribute prediction and retrieval.}
\Cref{fig:vis-qua-bad-attr} shows the main failure modes that remain after our model produces otherwise accurate 3-D bounding-box layouts.
Although the boxes fit the room well, the predicted object attributes—especially orientation—can be wrong. In the top-left scene and the two bottom scenes, beds or tables are oriented unnaturally after retrieval. At the room edges, some objects slide partly outside the plan, and in the top-right example, an L-shaped sofa is distorted because its width and length were swapped during retrieval. These cases show that attribute prediction, rather than box placement, is now the primary source of error.

\subsection{Full results for per-masktype model}
\label{sec:supp-expr-per-archtype}

\mypara{Evaluation by room type.} In \cref{tab:quant-distmatch-full,tab:quant-plausibility-full}, we present a comparison of \diffuscene, \midiff to our main \semdiff model (trained for each different type of conditioning) with breakdown for each room type.
Across the three room types, our \semdiff outperforms the other methods for almost all metrics, with \diffuscene having the best CKL for dining rooms, and \midiff having the best CKL for living rooms when conditioning on no arch or just the floor mask. 

\mypara{Statistics of generated scenes.}
As the plausibility metrics (e.g., out-of-bounds, collisions, navigability) are affected by the number of objects in the scene, we provide in \cref{tab:objstats} the average number of objects in scenes generated by different methods (with architecture conditioning) across different room types.  We see that on average, \diffuscene has the highest number of objects, and that our \semdiff has a comparable number of objects to \midiff.  Even with a similar number of objects, our \semdiff generated scenes with considerably less out-of-bounds and collisions than the other methods.

\subsection{Direct evaluation on generated layout}
\label{sec:supp-expr-layout}

One of the most crucial aspects of indoor scene synthesis is creating a plausible room layout. Therefore, we evaluate layouts directly to better compare model performance. \Cref{tab:quant-bbox} presents results obtained by rendering top-down views from each method’s pure generated layout with 3D object bboxes (prior to object retrieval) and computing layout metrics. These results show that our \semdiff significantly outperforms other methods in generating coherent and realistic layouts.

\subsection{Experiments with mixed-condition model}
\label{sec:supp-mix}

\begin{figure}[ht]
\includegraphics[width=\linewidth]{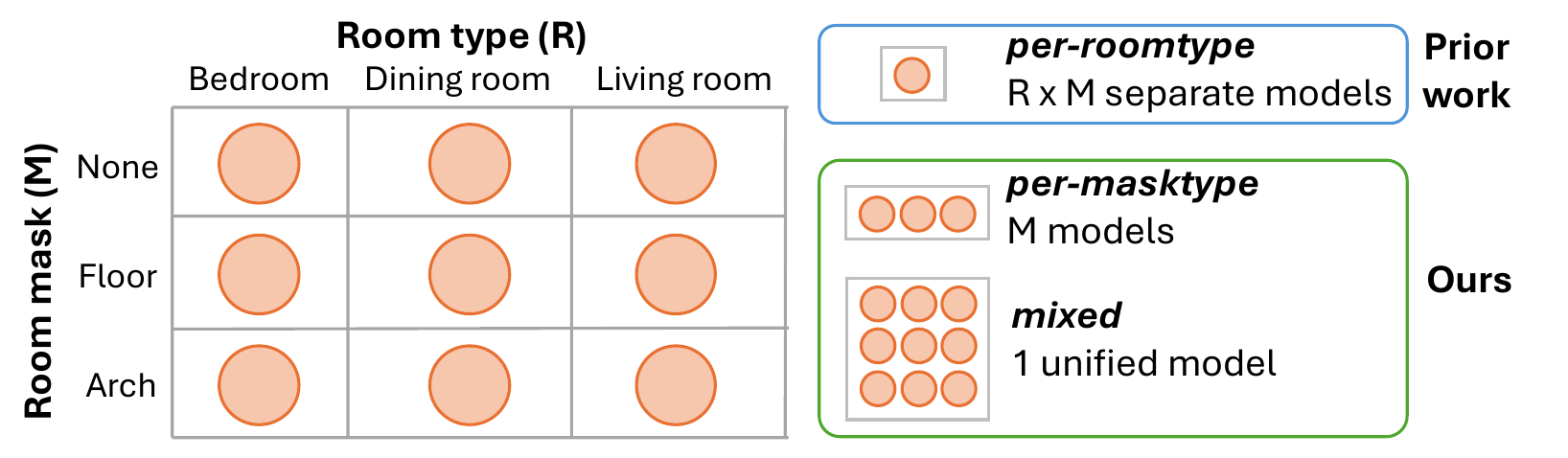}
\caption{
Prior work~\cite{paschalidou2021atiss,tang2023diffuscene} typically trained a separate model for each room type and conditioning type.  In this work, we explore unified models that combine all room types.  Most experiments in the main paper are conducted with per-masktype training, where we train a model for each different type of room mask conditioning (none, floor, arch).  We also investigate mixed training with a single unified model.
}
\vspace{-8pt}
\label{fig:model-types}
\end{figure}

We propose a mixed condition model that unifies generation with different room mask conditioning: no room mask (none), floor mask, and arch mask within a single framework. 
During training, each sample randomly selects one of the three condition types with equal probability (1/3). 
A condition type indicator (0: none, 1: floor, 2: arch) is introduced as an additional input.
It is processed through an embedding layer and MLP to obtain a condition type embedding, which is added to the timestep embedding—similar to the integration of room type embeddings. 
During inference, the model receives the desired condition type and its corresponding mask: no room mask uses an all-zero mask, the floor mask uses a binary mask (0 for void, 1 for floor), and the arch mask uses a categorical mask (0 for void, 1 for floor, 2 for door, 3 for window). 
This modification enables a single model to handle multiple room types and varying room mask conditional constraints flexibly.

As shown in \cref{tab:quant-distmatch-mix}, the mixed condition model achieves comparable performance to the per-masktype single-condition model with architecture conditioning, with only slight differences in FID, KID, SCA, and CKL metrics. 
Notably, the mixed condition model consistently improves results under the no room mask (none) and floor conditions across all room types. 
For instance, in the \texttt{None} condition, the mixed model reduces the overall average FID from 93.93 to 72.51, KID from 10.72 to 11.27, and CKL from 17.21 to 5.37, while increasing SCA from 96.76$\%$ to 92.81$\%$. 
For the \texttt{Floor} condition, the mixed model also outperforms the per-masktype single-condition model in FID, KID, and CKL. 
These results indicate that the mixed condition model improves performance under unconditional and floor constraints, while maintaining comparable performance for architecture conditioning.

\cref{tab:quant-plausibility-mix} presents the comparison of physical plausibility across different condition settings, evaluating both the mixed-condition model and the single-condition models.
For the unconditional case, the results show very low OOB values, which is expected since the floor and objects are generated jointly, ensuring boundary consistency. 
Across floor and arch conditions, the mixed-condition model achieves performance comparable to the single-condition models, with similar or slightly lower OOB and COL values and consistently high NAV scores. 
This indicates that the mixed model can generalize well across different condition types without sacrificing spatial plausibility, while still benefiting from a unified training scheme.

\subsection{Experiments with per-roomtype models}
\label{sec:supp-expr-per-room-model}

As most prior work on scene generation that learn the distribution of layouts from data are trained so that there is a separate model for each room type (and for each input condition type), we compare the performance of the different methods under this setting. 

Results \cref{tab:quant-distmatch-single-arch,tab:quant-plausibility-single-arch} show that \semdiff offers the best balance of adherence to the architecture plan and realism. 
At the scene level, it has lower out-of-boundary (\oobscene) errors ($14.3\%$ in bedrooms) and keeps them close to \midiff for the other room types, while at the object level it gives the lowest \oobobj in every room (e.g., $4.5\%$ vs.\ $11.7\%$ for \midiff in bedrooms and $10.0\%$ vs. $17.9\%$ in living rooms). 
Its overall CKL drops to $16.6$, far below \midiff’s $44.2$ and \diffuscene’s $82.4$, indicating the closest match to the ground-truth distribution. 
Collision rates are lowest in dining ($20.2\%$) and remain competitive in the other rooms, while navigability stays above $94\%$. 
Although \diffuscene and \midiff attain lower FID/KID in some cases, \semdiff combines comparable perceptual quality with superior architectural compliance and scene plausibility.

Note that \midiff{}’s checkpoint-selection procedure—picking the model with the lowest validation loss—works for floor-plan conditioning but breaks down when we adapt the code to architecture-plan conditioning for bedroom. 
The ``best" bedroom checkpoint consistently generates scenes filled almost exclusively with kid beds, many placed outside the room boundary.
To obtain evaluable bedroom results without otherwise altering MiDiffusion, we therefore inspected several intermediate checkpoints and chose the one that (i) positions furniture at a plausible distance from the room centre and (ii) produces a more varied, realistic mix of bedroom items.  
Checkpoints for all other room types remain unchanged, preserving alignment with the original \midiff protocol.

\subsection{Comparision with PhyScene}
\label{sec:supp-physcene}

We also provide a comparison of our \semdiff against PhyScene in \cref{tab:quant-physcene}.  However, as the PhyScene training code is not publicly available, we are unable to adapt it to our unified model setting.
For comparison, we take the checkpoint provided by PhyScene for living room (the only checkpoint that is publicly available), and we train our model under that setting (single-room type, floor-plan conditioning). 
Compared with PhyScene, our \semdiff consistently produces layouts that were more physically plausible with less collisions between objects, and less out-of-bounds errors.   

\begin{figure}
\centering
\setkeys{Gin}{width=0.95\linewidth}
\begin{tabularx}{\linewidth}{@{} Y | Y  Y  Y  @{}}
\toprule
\small{Arch} & \includegraphics[trim={30px 40px 30px 35px},clip]
{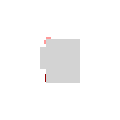} & \includegraphics[trim={30px 40px 30px 35px},clip]
{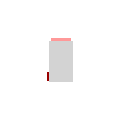} & \includegraphics[trim={20px 40px 20px 35px},clip]
{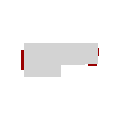}\\
\midrule
\small{Generated results} & \includegraphics[trim={0px 10px 0px 15px},clip]{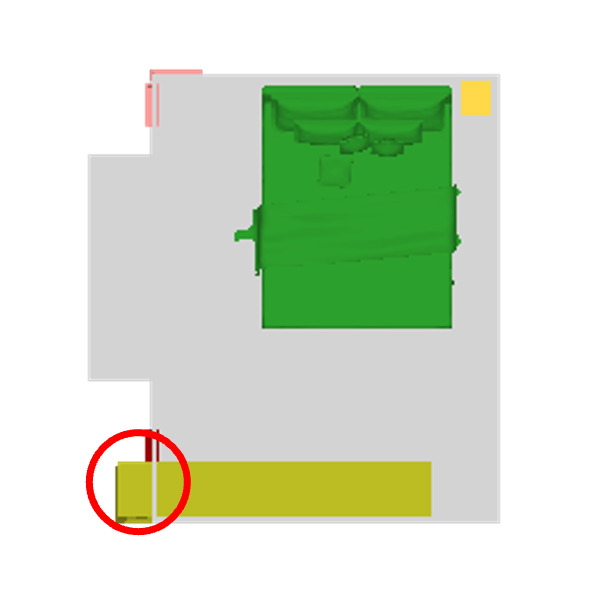} & 
\includegraphics[trim={0px 15px 0px 20px},clip]{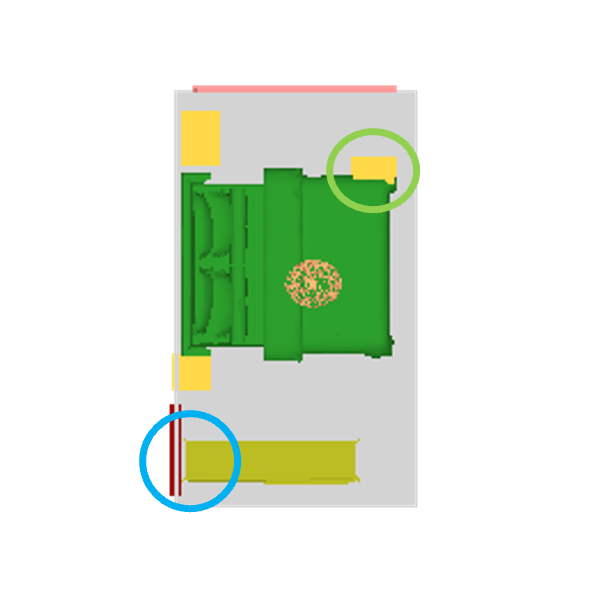} & 
\includegraphics[trim={0px 30px 0px 25px},clip]{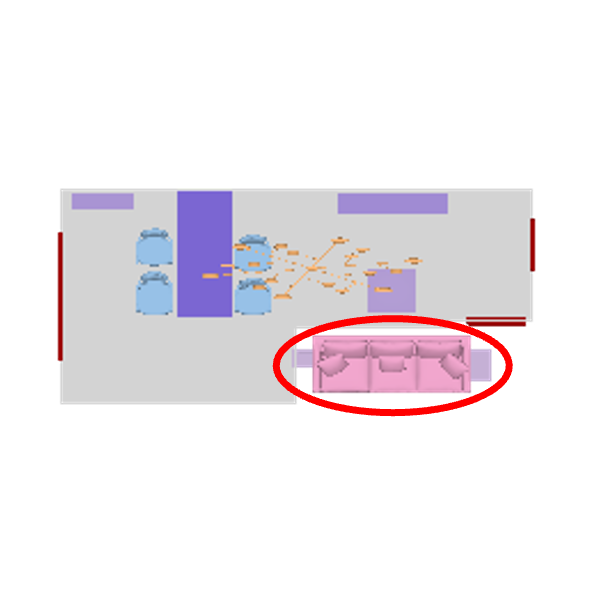}  \\
\bottomrule
\end{tabularx}
\vspace{-8pt}
\caption{Examples of failure cases from \semdiff-generated scenes with archplan conditioning.
}

\vspace{-1em}

\label{fig:vis-failure-cases}
\end{figure}

\subsection{Discussion of failure cases}
\label{sec:supp-failure-cases}
\Cref{fig:vis-failure-cases} shows several cases where there are issues with scenes generated by our \semdiff. First, objects can be placed outside the room boundaries (first and third images), primarily due to inaccurate attribute predictions (e.g. incorrect object size or orientation). Second, generated objects sometimes block doors, likely because the model lacks explicit constraints or losses to penalize placements near doors. Lastly, redundant objects may appear (middle image, extra nightstand) due to residual noise in the semantic map generated by the diffusion model, which the subsequent attribute model does not effectively eliminate.

\subsection{Additional qualitative results}
\label{sec:supp-expr-qualitative-results}

\begin{figure*}[ht]
\includegraphics[width=\textwidth]{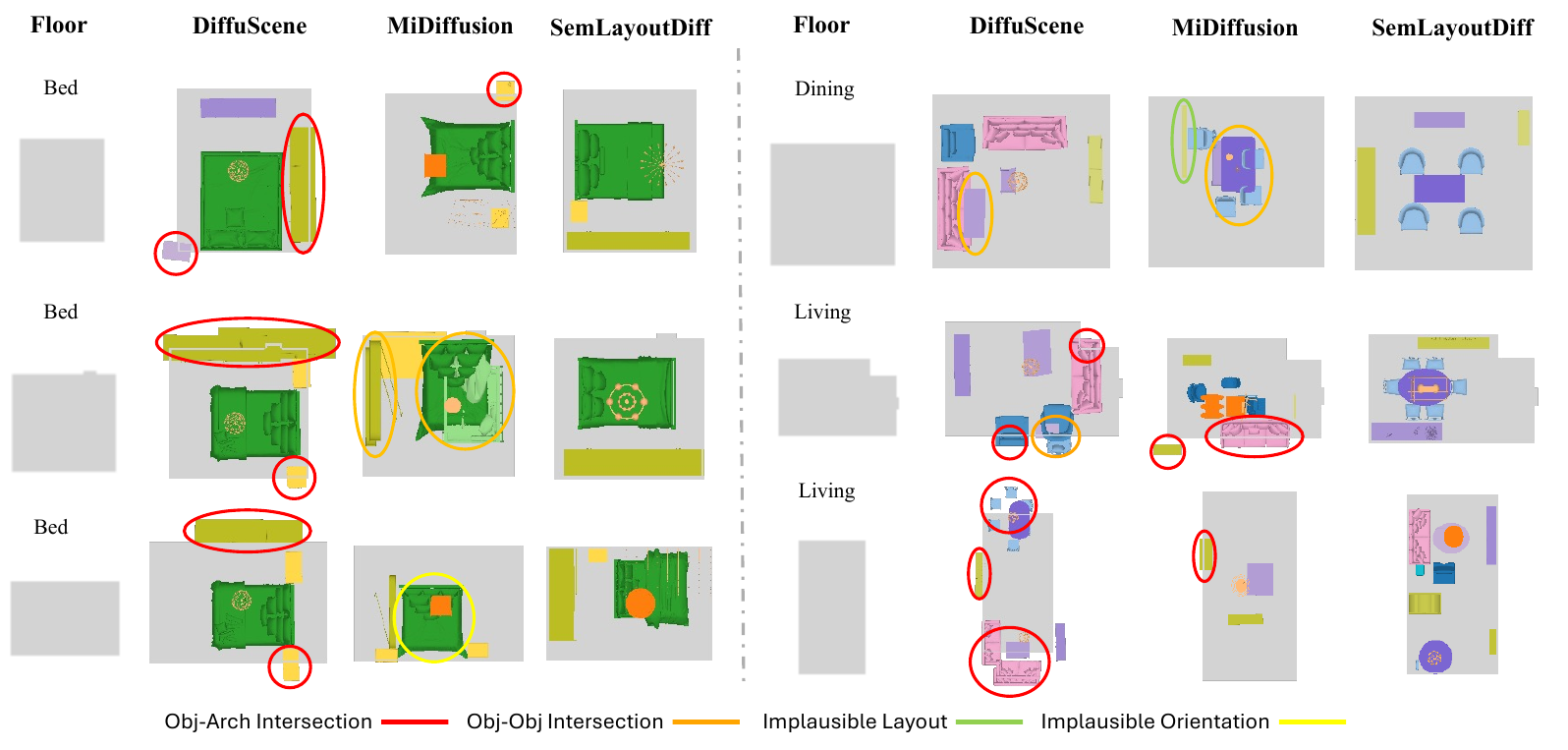}
\caption{Comparison of generated scenes using different unified methods with floor conditioning. The left four columns show bedroom scenes; the right four show living and dining room scenes.
}
\label{fig:vis-qua-floor}
\end{figure*}

\begin{figure*}[ht]
\includegraphics[width=\textwidth]{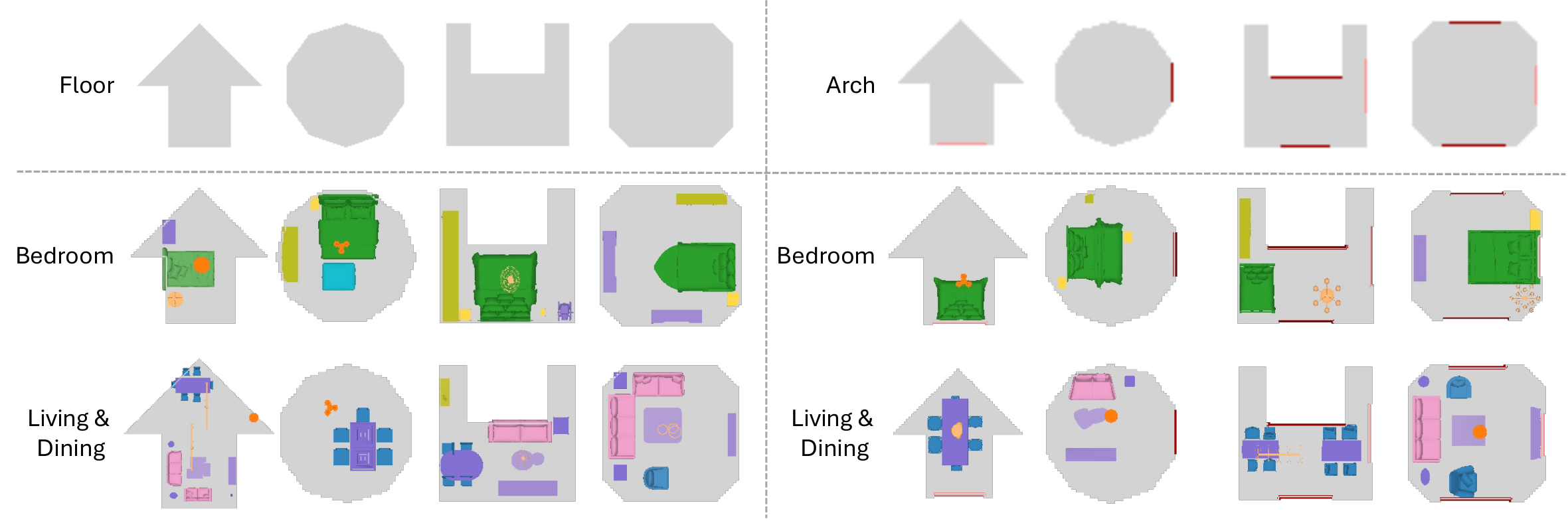}
\caption{Examples of generated scenes using our \semdiff mixed-condition model under custom floor and arch conditions.
}
\label{fig:vis-qua-custom}
\end{figure*}

\mypara{Floor conditioning results.} We provide floor conditioning results comparison with the unified models in \cref{fig:vis-qua-floor}.
We showcase our model's ability to fit to unusual custom floor masks in \cref{fig:vis-qua-custom}.

\begin{figure*}[ht]
\includegraphics[trim={0px 0px 10px 5px},width=0.9\textwidth]{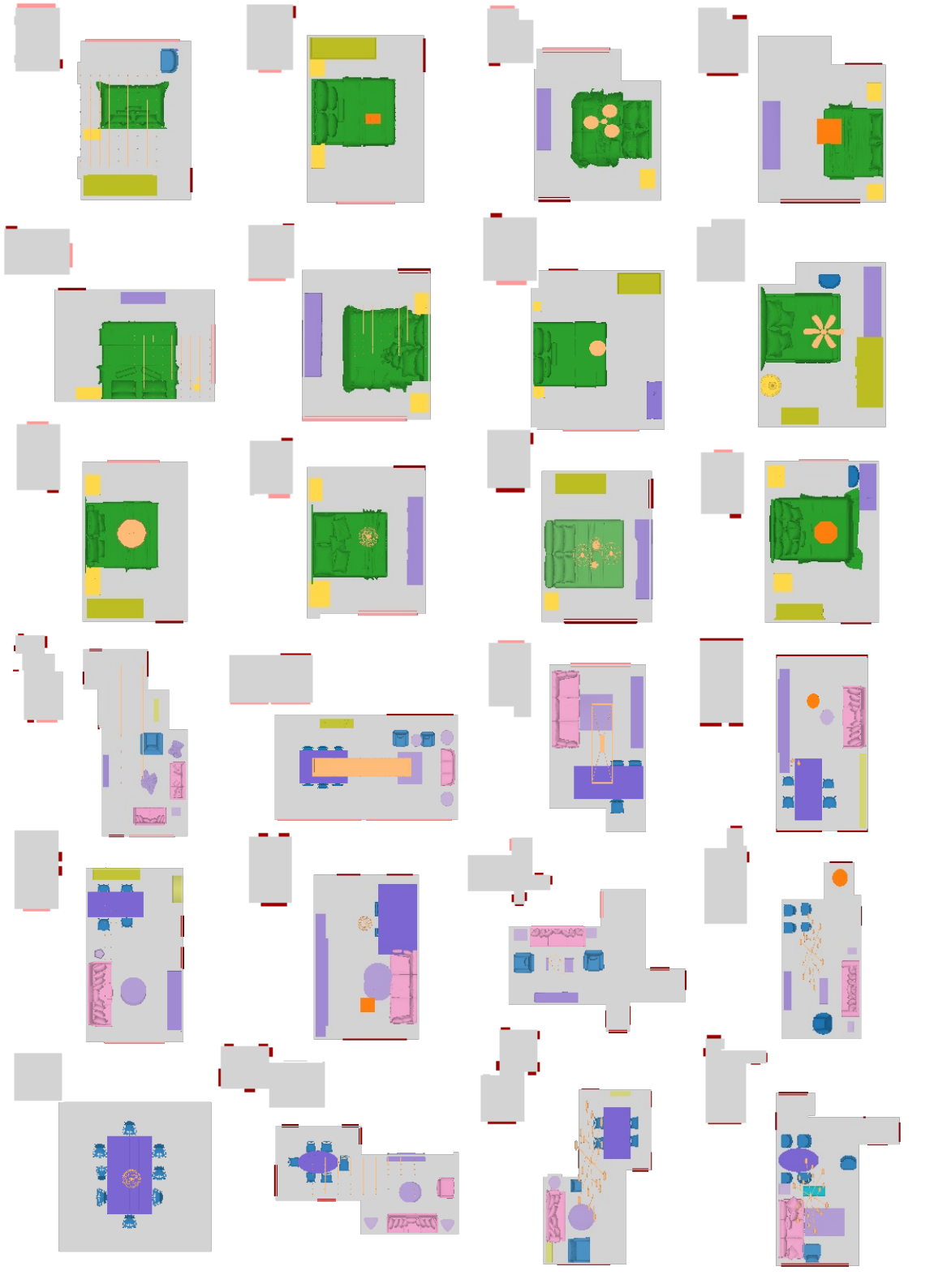}
\centering
\caption{
Examples of architecture-conditioned generated scenes using our \semdiff with orthographic rendering as used for evaluation. The condition arch mask is shown at the top-left of each result.
}
\vspace{-8pt}
\label{fig:vis-qua-evaluation-rendering}
\end{figure*}

\begin{figure*}[ht]
\includegraphics[width=\textwidth]{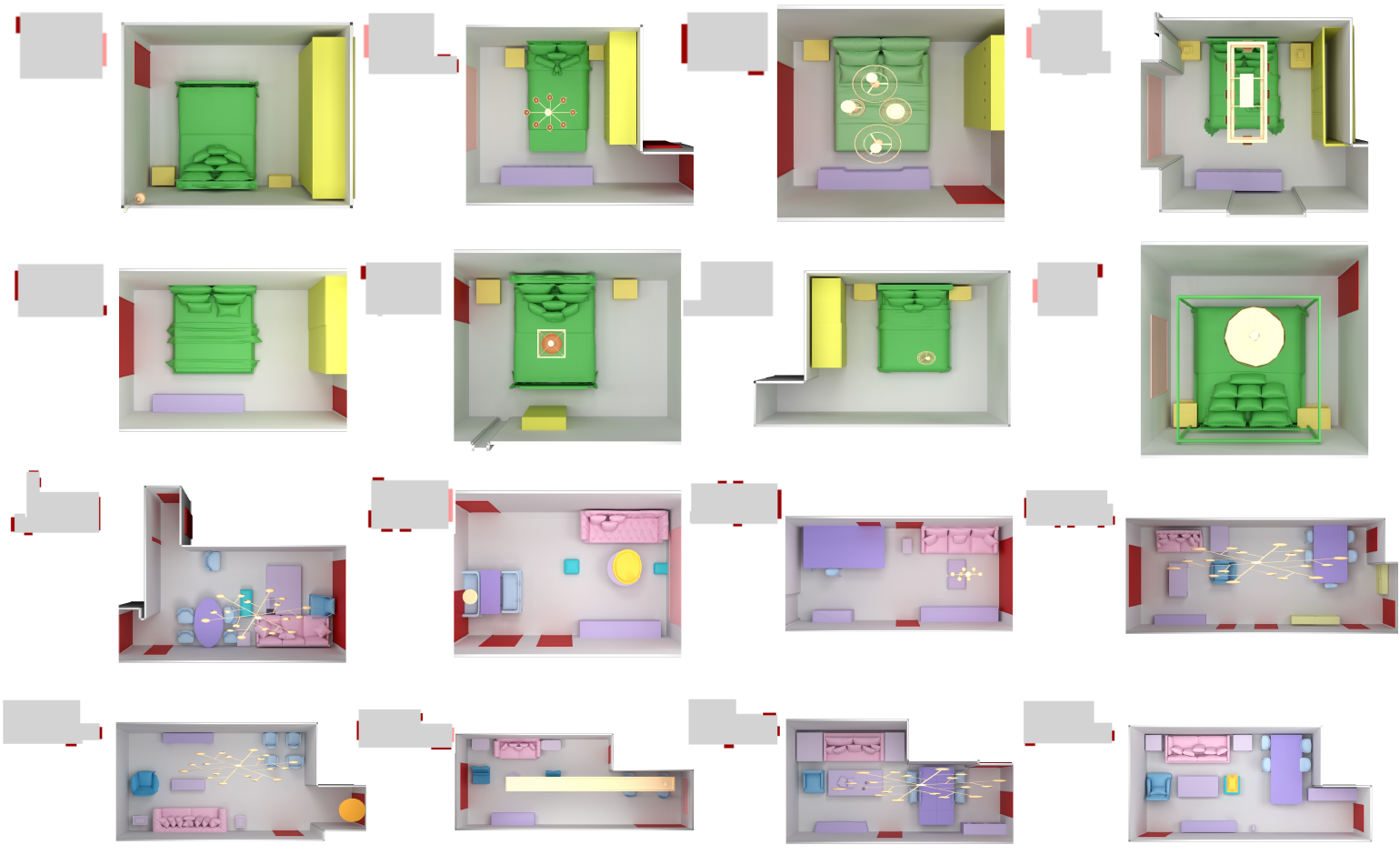}
\caption{
Examples of architecture-conditioned generated scenes using our \semdiff from top-down view with Blender rendering. The top left of each room is the provided arch mask condition.
}
\vspace{-8pt}
\label{fig:vis-qua-blender-semantic-coloring}
\end{figure*}

\begin{figure*}[ht]
\centering
\includegraphics[width=0.8\textwidth]{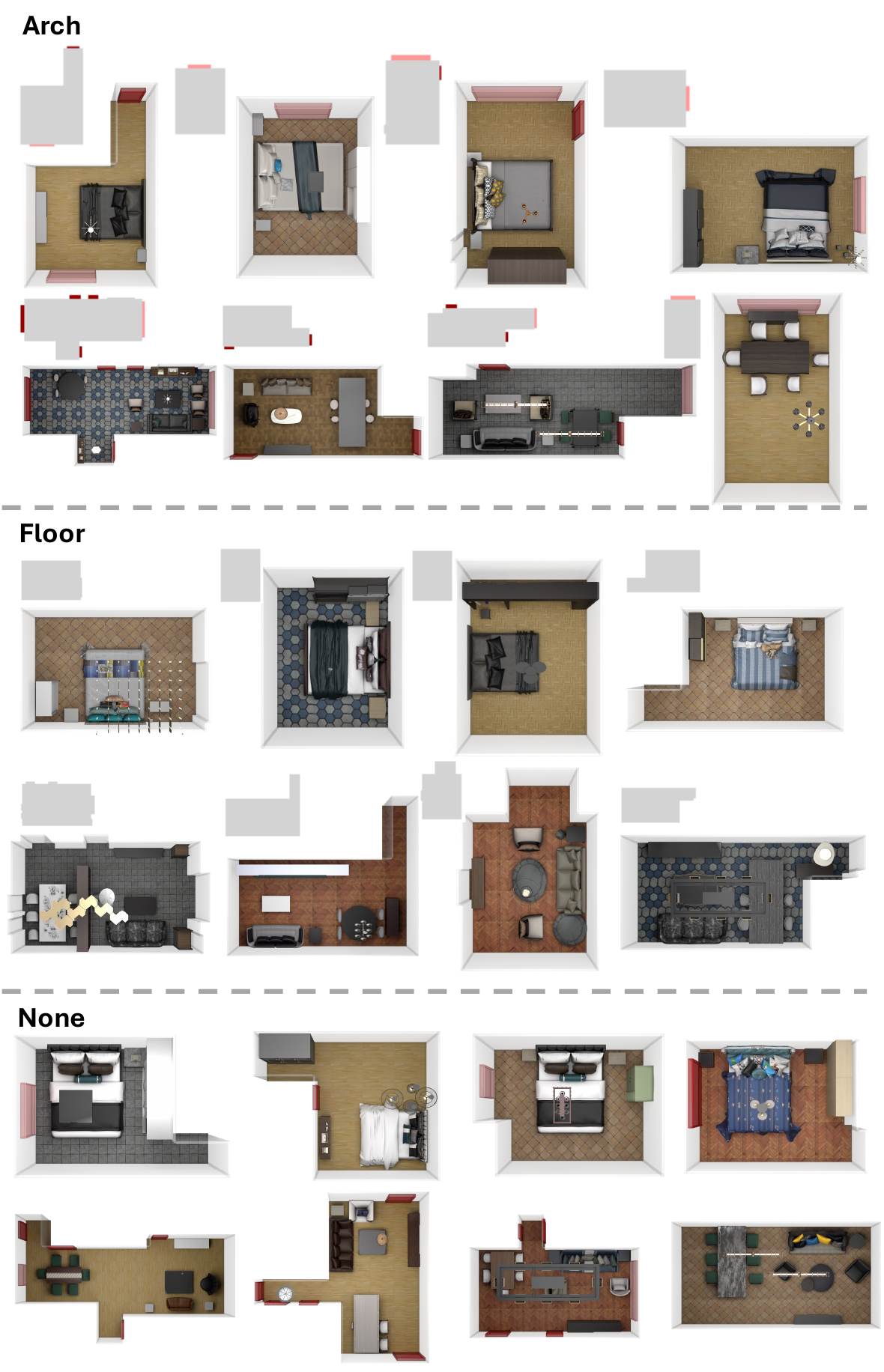}
\caption{
Examples of generated scenes using our \semdiff mixed-condition model under different condition types from top-down view with Blender rendering.
}
\vspace{-8pt}
\label{fig:vis-mix-texture}
\end{figure*}

\mypara{Additional renderings.} We provide additional renderings of scenes generated by \semdiff under architectural plan conditioning.
\Cref{fig:vis-qua-custom} also shows our model's ability to fit to unusual custom arch masks.
\Cref{fig:vis-qua-evaluation-rendering} shows top-down orthographic renderings using the same rendering setup as for quantitative evaluation. Additionally, we provide more top-down perspective renderings in both semantic colors (\cref{fig:vis-qua-blender-semantic-coloring}) and textured (\cref{fig:vis-mix-texture}) to better show the generated 3D layouts.

\end{document}